\RequirePackage{lineno} 

\documentclass[twocolumn,showpacs,preprintnumbers,amsmath,amssymb,nofootinbib]{revtex4}

\usepackage{graphicx}
\usepackage{dcolumn}
\usepackage{bm}

\def\bea{\begin{eqnarray}}
\def\eea{\end{eqnarray}}

\def\pp{\mbox{$p$-$p$}}
\def\auau{\mbox{Au-Au}}

\def\aa{\mbox{A-A}}
\def\nn{\mbox{$N$-$N$}}

\def\ee{\mbox{$e^+$-$e^-$}}

\def\qqbar{\mbox{$q$-$\bar q$}}
\def\pt{$p_t$}
\def\yt{$y_t$}

\def\v2{$v_2$}

\usepackage{epstopdf}
\begin{document} 

\setlength{\pdfpagewidth}{8.5in}
\setlength{\pdfpageheight}{11in}

\setpagewiselinenumbers
\modulolinenumbers[5]

\preprint{Version 4.2}

\title{
Transverse-rapidity $\bf y_t$ dependence of the nonjet azimuth quadrupole\\ from 62 and 200 GeV Au-Au collisions
}

\author{David T.\ Kettler}\affiliation{CENPA 354290, University of Washington, Seattle, Washington 98195}
\author{Duncan J.\ Prindle}\affiliation{CENPA 354290, University of Washington, Seattle, Washington 98195}
\author{Thomas A.\ Trainor}\affiliation{CENPA 354290, University of Washington, Seattle, Washington 98195}


\date{\today}

\begin{abstract}
Previous measurements of a quadrupole component of azimuth correlations denoted by symbol $v_2$ have been interpreted to represent elliptic flow, a hydrodynamic phenomenon conjectured to play a major role in noncentral nucleus-nucleus collisions. $v_2$ measurements provide the main support for conclusions that a ``perfect liquid'' is formed in heavy ion collisions at the Relativistic Heavy Ion Collider (RHIC). However, conventional $v_2$ methods based on one-dimensional (1D) azimuth correlations give inconsistent results and may include a jet contribution. In some cases the data trends appear to be inconsistent with hydrodynamic interpretations.
In this study we distinguish several components of 2D angular correlations and isolate a {\em nonjet} (NJ) azimuth quadrupole denoted by $v_2\{\text{2D}\}$. We establish systematic variations of the NJ quadrupole on $y_t$, centrality and collision energy.  We adopt transverse rapidity $y_t$ as both a velocity measure and as a logarithmic alternative to transverse momentum $p_t$. Based on NJ quadrupole trends we derive a {completely factorized} universal parameterization of quantity $v_2\{\text{2D}\}(y_t,b,\sqrt{s_{NN}})$ which describes the centrality, $y_t$ and  energy dependence.
From $y_t$-differential $v_2(y_t)$ data we isolate a {\em quadrupole spectrum} and infer a {\em quadrupole source boost} having unexpected properties.
NJ quadrupole $v_2$ trends obtained with 2D model fits are remarkably simple. The centrality trend appear to be uncorrelated with a {\em sharp transition} in jet-related structure that may indicate rapid change of Au-Au medium properties. The lack of correspondence suggests that the NJ quadrupole may be insensitive to such a medium. Several quadrupole trends have interesting implications for hydro interpretations.
\end{abstract}

\pacs{25.75.Ag, 25.75.Bh, 25.75.Ld, 25.75.Nq}

\maketitle

 \section{Introduction}

Measurements of the quadrupole component of azimuth $\phi$ correlations from RHIC heavy ion collisions in the form $v_2 = \langle \cos(2\, \phi)\rangle$ relative to estimates of the \aa\ reaction-plane angle are conventionally interpreted to represent {\em elliptic flow}, a conjectured hydrodynamic response to pressure gradients in the initial collision system corresponding to the overlap eccentricity of colliding nuclei~\cite{ollitrault}. In  a hydrodynamic (hydro) context~\cite{hydro1,hydro2,hydro3} inferred large elliptic flow values combined with other measurements are interpreted to imply rapid thermalization and production of a QCD medium with large energy density and small viscosity described as a ``perfect liquid''~\cite{perfliq,qgp1,qgp2}. 

However, questions persist concerning $v_2$ measurements and interpretations. Conventional 1D $v_2$ methods~\cite{poskvol,2004} may not distinguish accurately between a {\em nonjet} (NJ) azimuth quadrupole ({\em cylindrical multipole} uniform on pseudorapidity $\eta$ over a significant interval near midrapidity) and certain jet-related angular correlations that vary strongly with $\eta$ near midrapidity~\cite{axialci,ptscale,edep,lepmini}.  The terms jet-related and nonjet are discussed in Sec.~\ref{nomen}.

In previous studies we introduced a {physical-model-independent method} to distinguish {\em geometrically} between  a {NJ} quadrupole and the quadrupole ($m=2$) Fourier component of jet-related angular correlations dominated by a 2D peak centered at the origin on $\eta$ and $\phi$ difference variables~\cite{davidhq,noelliptic}.
The notation $v_2\{\text{2D}\}$ distinguishes the quadrupole component derived from model fits to 2D histograms from $v_2\{\text{method}\}$ data inferred with conventional 1D methods. We observed that $p_t$-integral $v_2\{\text{2D}\}(b,\sqrt{s_{NN}})$ data follow simple trends described by a few parameters over a broad range of centrality and collision energy $\sqrt{s_{NN}}$ above 13 GeV. The trends factorize, each factor described by a simple function. That analysis was complementary to an analysis reported in Ref.~\cite{anomalous} focusing on jet-related structure.

In the present study we extend the NJ quadrupole program to measurements of $y_t$-differential $v_2\{\text{2D}\}(y_t,b,\sqrt{s_{NN}})$ also derived from 2D model fits. As an alternative to transverse momentum \pt\ we introduce transverse rapidity \yt\ as a logarithmic variable compatible with relativistic boost measurements. From $y_t$-differential $v_2\{\text{2D}\}$ data it is possible to infer a {\em quadrupole source boost} distribution common to hadrons of several species~\cite{davidhq2,quadspec}. In this study we determine the \auau\ centrality dependence of the quadrupole source boost. We also infer a corresponding {\em quadrupole spectrum} common to several hadron species~\cite{quadspec} and substantially different from the spectrum for most final-state hadrons. Those results offer new insights into possible mechanisms for the NJ azimuth quadrupole. 

This paper is arranged as follows: 
In Sec.~\ref{genan} we introduce some general correlation analysis methods. 
In Secs.~\ref{nongraph} and \ref{modelfit} we describe two alternative methods for estimating azimuth quadrupole components of angular correlations. 
In Sec.~\ref{2dang} we introduce measured  $y_t$-differential 2D angular autocorrelations (histograms) derived from particle data. 
In Sec.~\ref{ytdiffsyst}  we review systematic model-parameter trends from model fits to the 2D data histograms. 
In Sec.~\ref{quadboostt} we define the quadrupole source boost and determine  its centrality dependence.
In Sec.~\ref{quadspecc} we extract quadrupole spectra and describe the centrality dependence.
In Sec.~\ref{v2ytparam} we derive a universal factorized parametrization of non-jet quantity $v_2\{\text{2D}\}(y_t,b,\sqrt{s_{NN}})$. 
In Sec.~\ref{uncert} we discuss systematic uncertainties, and in Sec.~\ref{methcomp} we present comparisons between quadrupole amplitudes derived from 2D fits to angular correlations and from other $v_2$ methods.
In Secs.~\ref{discc} and~\ref{summ} we present discussion and summary.

{\tt 


}

 \section{General Analysis methods} \label{genan}

In this study we report measurements of  $y_t$-differential $V_2^2\{\text{2D}\}(y_t,b)$ nonjet azimuth {\em power-spectrum} elements derived from model fits to 2D angular correlations. Transverse rapidity $y_t$ (defined below) serves as a logarithmic measure of transverse momentum $p_t$.

\subsection{Kinematic measures and spaces}

\aa\ collisions with impact parameter $b$ produce final-state hadrons in cylindrical 3D momentum space $(p_t,\eta,\phi)$, where $p_t$ is transverse momentum, $\eta$ is pseudorapidity and $\phi$ is azimuth angle. Transverse mass is $m_t = \sqrt{p_t^2 + m_h^2}$ with hadron mass $m_h$. Pseudorapidity is defined by $\eta = -\ln[\tan(\theta/2)] $ ($\theta$ is polar angle relative to collision axis $z$), and $\eta \approx \cos(\theta)$ near $\eta = 0$. Transverse rapidity is defined by $y_t = \ln[(m_t + p_t) / m_h]$. For identified hadrons the proper hadron mass is used and $y_t$ is then a velocity measure appropriate to test flow conjectures. For unidentified hadrons $y_t$ with pion mass assumed (about 80\% of hadrons) serves as a logarithmic measure of $p_t$, and the pion mass regularizes the logarithmic trend for small values of $p_t$. The STAR TPC acceptance $p_t > 0.15$ GeV/c corresponds to $y_t > 1$.


Two-particle correlations are structures in the pair density on 6D momentum space $(y_{t1},\eta_1,\phi_1,y_{t2},\eta_2,\phi_2)$. Angular correlations can be measured on subspace $(\eta_1,\eta_2,\phi_1,\phi_2)$ given some conditions on transverse momentum $(p_{t1},p_{t1})$ or transverse rapidity $(y_{t1},y_{t2})$. We can integrate over all $y_t$ ($y_t$-integral analysis) or define conditions on two-particle  $(y_{t1},y_{t2})$ ($y_t$-differential analysis). Alternatively, we can integrate over some part of the angular acceptance (angular acceptance conditions) to study conditional correlations on $(y_{t1},y_{t2})$~\cite{porter2,porter3,pptrig1,pptrig2}.

An {\em autocorrelation} on angular subspace $(x_1,x_2)$ is derived by averaging pair density $\rho(x_1,x_2)$ along diagonals on $(x_1,x_2)$ parallel to the sum axis $x_\Sigma = x_1 + x_2$~\cite{inverse}. The averaged pair density $\rho(x_\Delta)$ on defined {\em difference variable} $x_\Delta = x_1 - x_2$ is then an autocorrelation~\cite{inverse}. For correlation structure approximately independent of $x_\Sigma$ over some limited acceptance $\Delta x$ (stationarity, typical over $2\pi$ azimuth and within some limited pseudorapidity interval $\Delta \eta$) angular correlations remain undistorted (no information is lost in the projection by averaging)~\cite{hadrogeom}.   Within the STAR TPC acceptance $\Delta \eta = 2, \Delta \phi = 2\pi$~\cite{starnim} 2D angular autocorrelations are lossless projections of 6D two-particle momentum space onto angle difference axes $(\eta_\Delta,\phi_\Delta)$~\cite{flowmeth}. 
The $\phi_\Delta$ axis is divided into {\em same-side} (SS, $|\phi_\Delta| < \pi/2$) and {\em away-side}  (AS, $\pi/2 < |\phi_\Delta| < \pi$) regions.

In the present analysis we impose conditions on the space $(y_{t1},y_{t2})$ to establish the full ($y_t,b,\sqrt{s_{NN}}$) systematics of angular correlations, emphasizing the NJ quadrupole obtained from 2D model fits via two model parameters: {\em per-particle} quadrupole amplitude $A_Q\{\text{2D}\}$ or {\em per-pair} amplitude $B_Q\{\text{2D}\}$ (terms defined below). 

\subsection{Correlation structure and interpretations} \label{nomen}

The 2D data histograms that form the basis for this study (within the STAR TPC acceptance) exhibit the same three dominant features from \pp\ to central \auau\ collisions: (a) a SS 2D peak centered at the origin on $(\eta_\Delta,\phi_\Delta)$, (b) an AS 1D peak centered at $\phi_\Delta = \pi$ and approximately uniform on $\eta_\Delta$, and (c) an azimuth quadrupole component uniform on $\eta_\Delta$. Those three elements are distinguished in all cases by model fits to 2D angular correlations on $\eta_\Delta$ and $\phi_\Delta$.  Component (a) is well-described by a SS 2D Gaussian in most cases. With increasing \aa\ centrality the SS peak is elongated on $\eta_\Delta$. If high-$p_t$ (trigger-associated) cuts are applied the SS peak may develop non-Gaussian tails on $\eta_\Delta$. Component (b) is well-described by a single AS dipole term.

In \pp\ and more-peripheral \auau\ collisions components (a) and (b) represent intrajet and interjet correlations respectively: Their amplitudes scale with the number of binary \nn\ collisions $N_{bin}$ as expected for dijets, and their forms are consistent with pQCD jet structure predicted by \textsc{pythia}~\cite{pythia} and \textsc{hijing}~\cite{hijing}.  They both retain the same forms and follow $N_{bin}$ scaling (\nn\ linear superposition) in \auau\ collisions up to 50\% centrality~\cite{anomalous}. 
Throughout that centrality interval it is therefore appropriate to refer to (a) and (b) as {\em jet-related} structures. In more-central \auau\ collisions (above a {\em sharp transition} near 50\% centrality~\cite{anomalous}) the SS 2D peak becomes increasingly elongated on $\eta_\Delta$, and the peak amplitude increases faster than $N_{bin}$ scaling. However, other features of the SS peak remain consistent with jet production~\cite{jetsyield}. The AS dipole amplitude closely follows the SS 2D peak amplitude and also remains consistent with jet expectations. Nevertheless, other (nonjet) interpretations have been proposed for those structures~\cite{gunther,luzum}.

Azimuth quadrupole component (c) appears to be uncorrelated with jet-related components (a) and (b), for example exhibiting a smooth centrality dependence with no evidence of the sharp transition~\cite{anomalous,noelliptic}. In that context it is appropriate to refer to (c) as the NJ quadrupole. The NJ quadrupole inferred from 2D model fits (represented by symbol $v_2\{\text{2D}\}$) is an isolated quadrupole structure uniform on $\eta$ near mid-rapidity with maxima at $0$ and $\pi$ on azimuth. Its form is then consistent with conventional expectations for ``elliptic flow'' if that physical mechanism is relevant. The SS 2D peak (a) projected onto 1D azimuth can be modeled as a narrow Gaussian with its own Fourier series representation~\cite{multipoles}. The SS peak Fourier terms then contribute to any 1D Fourier description of all correlation structure combined. Resulting series terms $v_m$ then include admixtures of elements (a), (b) and (c). In this study we show that multiple (physical) contributions to such a 1D Fourier series can be distinguished. We therefore refer to NJ quadrupole (c) as the object of the present study distinct from a jet-related quadrupole derived from (a) which may be a source of systematic error for 1D $v_2$ analysis~\cite{gluequad}. 



\subsection{Joint and marginal distributions on $\bf (y_{t1},y_{t2})$}

\begin{figure}[h]
  \includegraphics[width=1.65in,height=1.65in]{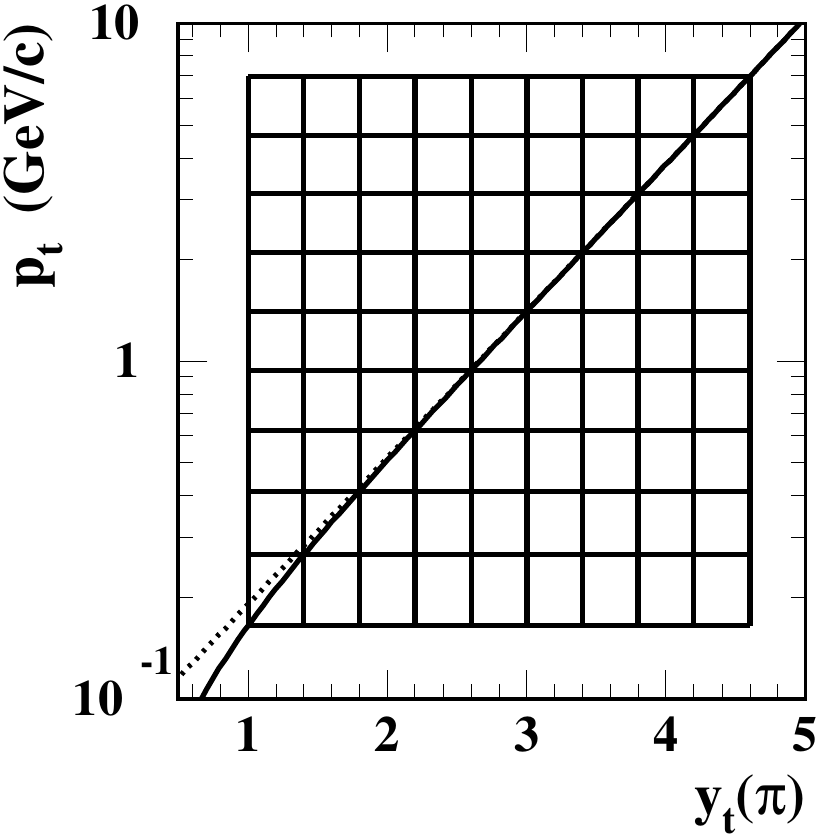}
  \includegraphics[width=1.65in,height=1.63in]{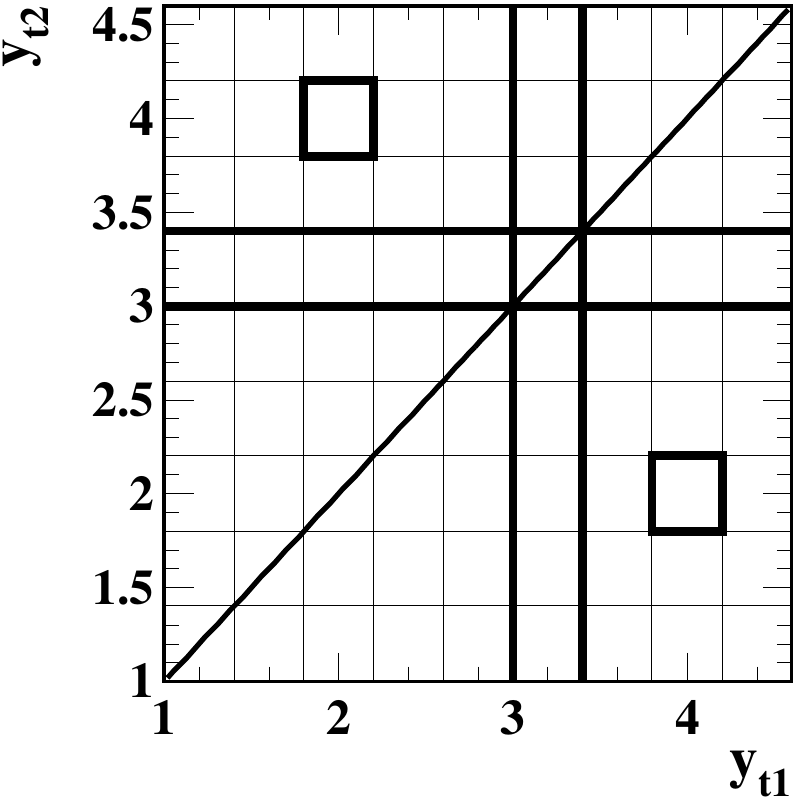}
\caption{\label{kine}
Left: The relation between transverse momentum $p_t$ and transverse rapidity $y_t(\pi)$ (assuming a pion mass). The grid shows the uniform $y_t$ bin system used for the present analysis. The dotted line provides a $\log(p_t)$ reference for comparison.
Right: The symmetrized two-particle space $(y_{t1},y_{t2})$. An element of the {\em joint} distribution on $(y_{t1},y_{t2})$ is shown by the bold squares. An element of the {\em marginal} distribution on $y_t$ is shown by the bold rectangles. 
} 
 \end{figure}

This analysis addresses angular correlation systematics corresponding to various cut conditions on space $ (y_{t1},y_{t2})$. Fig.~\ref{kine} (left panel) shows the relation between transverse momentum $p_t$ and transverse rapidity $y_t$. $y_t \approx 1$ ($p_t \approx 0.15$ GeV/c) is the typical lower $p_t$ bound defining the TPC acceptance for spectra and correlations. The dotted line $ (m_\pi / 2) \exp(y_t) \approx p_t $ demonstrates that $y_t$ accurately represents $\log(p_t)$ over the TPC $p_t$ acceptance. The close correspondence down to the $p_t$ acceptance limit arises because the pion mass is $\approx 0.1$ GeV/c. The grid illustrates the cut system for this analysis, nine bins on $y_t$ with fixed width $\delta y_t = 0.4$.  The $p_t$ interval covered by the analysis is [0.16,7] GeV/c. 

Figure~\ref{kine} (right panel) shows the binning on $(y_{t1},y_{t2})$ for the $y_t$-differential analysis. The binning system is symmetric about the diagonal. The bold squares illustrate an element of the  {\em joint} distribution on $(y_{t1},y_{t2})$. By integrating over one axis we obtain the 1D projection or {\em marginal} distribution on $y_t$ represented by the bold rectangles. The marginal format is the basis for the $y_t$-differential part of this analysis. Since the cut system (equal $y_t$ bins) and other aspects of this analysis are based on $y_t$ we prefer that quantity in the text, with occasional references to specific $p_t$ values. 

\subsection{Single-particle and correlated-pair measures} \label{spyield}

The single-charged-particle (SP) $y_t$ spectrum is  $\rho_0(y_t,b) = d^2 n_{ch}/ y_t dy_t 2\pi d\eta$ (azimuth averaged). The $y_t$-integral angular density is $\rho_0(b) = \int dy_t y_t \rho_0(y_t,b) \approx n_{ch} / 2\pi \Delta \eta$  averaged over acceptance $\Delta \eta$. 
The two-component particle-yield parametrization~\cite{kn} as applied in this analysis is $\rho_0(b) = \rho_{pp} (N_{part}/2)[1 + x (\nu - 1)] $ where $2\pi \rho_{pp}$ and $x$ are 2.5 and 0.095 respectively for more-central 200 GeV \auau\ collisions.
Glauber-model centrality parameters $N_{part}/2$ and $\nu$ are defined below.

$\rho(\vec p_{1},\vec p_{2})$ is the basic pair density on 6D pair momentum space.  The event-averaged pair density  $\rho_{sib}$ derived from sibling pairs (pairs drawn from single events) includes the correlation structures to be measured. $\rho_{mix}$ is the density of mixed pairs drawn from different but similar events. $\rho_{ref}$ denotes a minimally-correlated reference-pair density derived from (a) a mixed-pair density or (b) a product of SP densities via a factorization assumption. On pair subspace $(y_{t1},y_{t2})$ the factorized joint reference is $\rho_{ref}(y_{t1},y_{t2},b) = \rho_0(y_{t1},b)\rho_0(y_{t2},b)$, the marginal reference is $\rho_{ref}(y_{t},b) =\rho_0(b)\rho_0(y_{t},b)$ and the $y_t$-integral reference is $\rho_{ref}(b) =\rho_0^2(b)$. 

Formation of autocorrelation histograms on difference variables $(\eta_\Delta,\phi_\Delta)$ projected from pair angle subspace $(\eta_1,\phi_1,\eta_2,\phi_2)$ has been described previously~\cite{inverse,anomalous}. Pair histograms so formed are approximately uniform on $\eta_\Delta$ and $\phi_\Delta$. Small deviations from uniformity represent correlations of interest.  Pair histograms formed by simple projection from $(\eta_1,\eta_2)$ (not autocorrelations) include a triangular pair acceptance on difference variable $\eta_\Delta$. 

Differential correlation structure is determined by comparing a sibling-pair density to a reference-pair density in the form of difference $\Delta \rho = \rho_{sib} - \rho_{ref}$ representing a correlated-pair density or {\em covariance} density. There are then two choices for a {\em relative} correlation measure: 

(a) {\em Per-particle} measure $\Delta \rho / \sqrt{\rho_{ref}}$ has the form of Pearson's normalized covariance wherein the numerator is a covariance and the denominator is the geometric mean of marginal variances. In the Poisson limit a marginal variance is given by $\sigma^2_n = \bar n \propto \rho_0$. Since $\rho_{ref} \approx \rho_0 \times \rho_0$ it follows that the geometric mean of variances is given by  $\sqrt{\rho_\text{ref}} \propto n_{ch}$ and the normalized covariance is  a {per-particle} correlation measure~\cite{inverse,ptscale,edep}. 

(b) {\em Per-pair} measure $\Delta \rho / \rho_{\text{ref}} $ {\em decreases} trivially with system size as $1 / n_{ch}$. That trend obscures smaller but physically-meaningful variations. The per-pair measure also tends to {\em increase} trivially as a function of  $y_t$ because the pair ratio includes the SP spectrum in its denominator. The dominant SP spectrum trend also obscures physically-meaningful correlation variations.

In a practical correlation analysis pair ratio $\Delta \rho / \rho_{ref} \rightarrow \Delta \rho / \rho_{mix}$ is first calculated directly to cancel particle-pair detector inefficiencies. The per-particle measure $\Delta \rho / \sqrt{\rho_{ref}}\equiv \sqrt{\rho_{ref}}\Delta \rho / \rho_{mix}$ is then obtained, where $\rho_{ref}$ is constructed from {\em corrected} SP spectra $\rho_0(y_t,b)$ and yields $\rho_0(b)$~\cite{anomalous}.

In this \yt-differential analysis we present per-pair quadrupole measurements based on 2D model fits to angular correlations. 
The basic measures are the Fourier components $V_m^2$ of $\Delta \rho(\phi_\Delta)$ emphasizing the quadrupole term $V_2^2$. 
Per-pair ratio $\Delta \rho / \rho_{mix}$ gives $B_Q\{\text{2D}\}(y_t,b) = v_2^2\{\text{2D}\}(y_t,b)$ directly comparable with published $v_2(p_t)$ data. Per-particle measure $\sqrt{\rho_{ref}}\Delta \rho / \rho_{mix}$ gives $A_Q\{\text{2D}\}(b) = \rho_0(b) v_2^2\{\text{2D}\}(b)$ exhibiting simple systematic trends on centrality and collision energy~\cite{davidhq,noelliptic}. Since $V_2^2(y_t,b) = \rho_0(b) \rho_0(y_t,b)(b) v_2^2(y_t,b)$ we have $\rho_0(b) A_Q(b) = V_2^2(b)$ and $\rho_0(b)\rho_0(y_t,b) B_Q(y_t,b) = V_2^2(y_t,b)$, defining per-particle $A_Q(b)$ and per-pair $B_Q(y_t,b)$ quadrupole measures. Per-particle and per-pair measures are thus exactly related. For reasons noted above $V_2^2$ and per-particle measure $A_Q$ are the bases for physical interpretations. Per-pair measure $B_Q$ from this $y_t$-differential study does not require corrected SP spectra $\rho_0(y_t,b)$ and provides direct comparison with published $v_2(p_t)$ data.



\subsection{\aa\ centrality measures}

\aa\ centrality is measured by matching the fractional cross section $\sigma / \sigma_0$ for some observed $n_{ch}$ to the fractional cross section derived from a Glauber Monte Carlo simulation. Glauber parameters $N_{part}/2$ (participant pairs), $N_{bin}$ (\nn\ binary collisions) and impact parameter $b$ are thereby related to $n_{ch}$ integrated within the TPC acceptance $|\eta| < 1$. The fractional impact parameter is defined by $b / b_0 \equiv \sqrt{\sigma / \sigma_0}$. Centrality measure $\nu \equiv 2\, N_{bin} / N_{part}$ estimates the mean number of \nn\ {encounters} per participant nucleon (mean projectile-nucleon path length across the collision-partner nucleus). We use the same Glauber parameters for all energies as purely geometrical measures (the 200 GeV \nn\ cross section $\sigma_{NN} = 42$ mb is assumed for all cases). 

 \begin{figure}[h]
  \includegraphics[width=1.65in,height=1.65in]{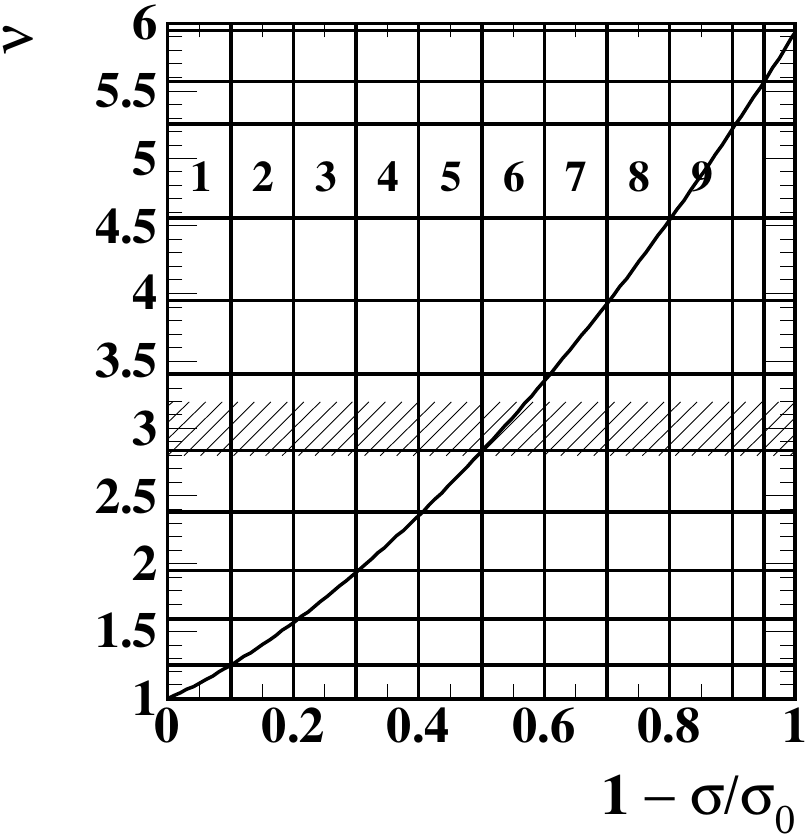}
  \includegraphics[width=1.65in,height=1.65in]{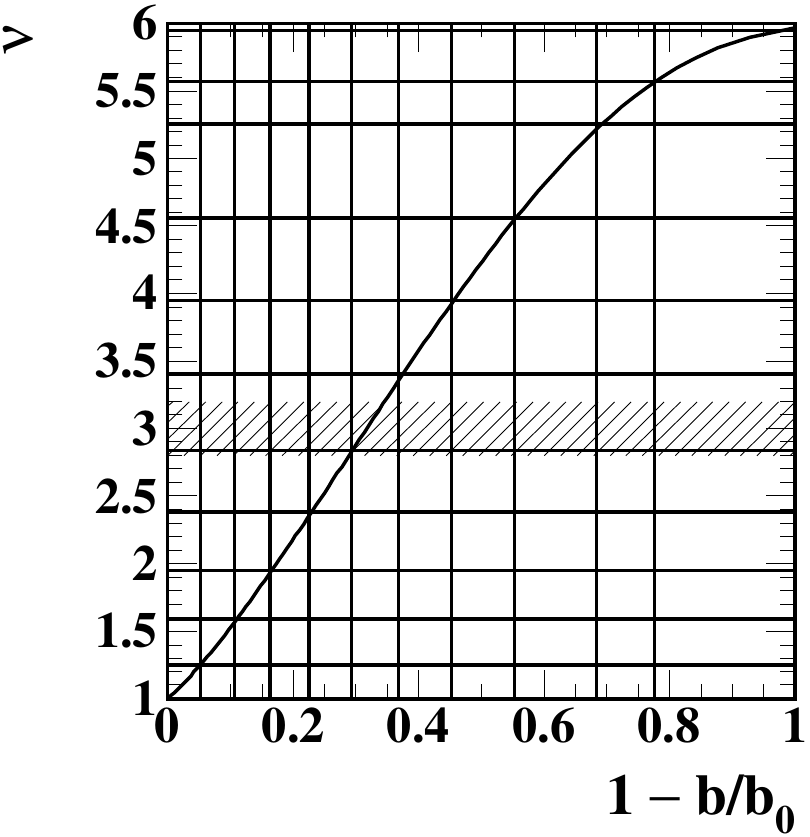}
\caption{\label{centro}
Left: Mean participant path length $\nu$ vs fractional cross-section measure $1 - \sigma / \sigma_0$. The grid shows the eleven centrality bins for this analysis. The hatched region indicates the ``sharp transition'' in jet-related correlation properties reported in Ref.~\cite{anomalous}.
Right: Path length $\nu$ vs fractional impact parameter $b / b_0 = \sqrt{\sigma / \sigma_0}$. For \auau\ collisions $b_0 \approx 14.7$ fm.
} 
 \end{figure}

Figure~\ref{centro} (left panel) shows participant path length $\nu$ vs fractional cross section in the form $1 - \sigma / \sigma_0$ inferred from a Glauber Monte Carlo. The hatched band, with position inferred from angular correlation data, represents a {\em sharp transition} in jet-related correlation systematics below which jet  correlations follow the \nn\ binary-collision scaling expected for linear superposition of \nn\ collisions (\aa\ transparency) and no jet modification~\cite{anomalous}. Fig.~\ref{centro} (right panel) shows $\nu$ vs fractional impact parameter as $1 - b / b_0$, where $b_0 \approx 14.7$ fm for \auau\ collisions. 

The grids in Fig.~\ref{centro} indicate the centrality bins for this analysis defined as follows: Uncorrected minimum-bias event samples are divided into 11 nominal centrality bins: nine $\approx10$\% bins from 100\% to 10\%, the last 10\% divided into two 5\% bins. The corrected centrality of each bin as modified by tracking and event vertex inefficiencies is determined 
by a running-integral procedure described  in Ref.~\cite{powerlaw}. Centralities from \nn\ collisions ($\nu \approx 1.25$) to central A-A ($b \approx 0$) are thereby determined to 2\%.

\subsection{\aa\ initial-state geometry measures} \label{geomsec}

Some features of the initial-state (IS) geometry of \aa\ collisions may influence collision dynamics.
 IS azimuth structure is conventionally modeled 
by a Glauber Monte Carlo. The participant-nucleon azimuth distribution can be described by an autocorrelation function on azimuth difference $\phi_\Delta$~\cite{inverse}. 
For non-central \aa\ collisions the autocorrelation includes (a) a few even-$m$ sinusoids dominated by $m=2$ (IS quadrupole) phase-correlated with vector impact parameter $\vec b$ (the eccentric \aa\ overlap region), (b) a uniform background and (c) a delta-function term $\propto N_{part}$ (self pairs) uncorrelated with $\vec b$. 

The Fourier transform of the IS azimuth autocorrelation is a power spectrum represented by eccentricity elements $E_m^2 = N_{part}^2 \epsilon_m^2$, with {\em per-pair} eccentricity measures~\cite{gluequad}
\bea \label{optical}
\epsilon_{m,MC}^2 &=& \epsilon_{m,opt}^2 + \sigma^2_{\epsilon_m} + \delta\epsilon_{m}^2~~\text{for $m$ even} \\ \nonumber
&=&  \delta\epsilon_{m}^2~~\text{for $m$ odd}.
\eea
$\sigma^2_{\epsilon_m}$ represents an eccentricity variance due to event-wise $b$ fluctuations.  
Eccentricities $\epsilon_{m,opt}^2$ ($m = 2,$ 4) represent the ``elliptical'' \aa\ overlap region for fixed $b$ and smooth matter distributions. 
The corresponding $m=2$ optical eccentricity for 200 GeV \auau\ is parametrized by~\cite{davidhq}
\bea \label{epsopt}
\epsilon_{{2,opt}} = \frac{1}{5.4} \left[\log_{10}\left(\frac{3\, N_{bin}}{2}\right)\right]^{0.96}  \left[\log_{10}\left(\frac{1136}{N_{bin}}\right)\right]^{0.78}\hspace{-.23in}.
\eea

Point-wise Monte Carlo random sampling generates a ``white-noise'' power spectrum $\delta\epsilon^2_m \propto 1/N_{part}$ (approximately uniform on $m$) corresponding to the self-pair contribution $\approx N_{part} \delta(\phi_\Delta)$ in the IS azimuth autocorrelation. For a stochastic process there should be no phase relation between noise amplitudes $\delta\epsilon^2_m$ and impact parameter $\vec b$. 
All higher $m$ are present in the IS Monte Carlo spectrum and {\em might} appear in the observed final state to some extent {\em if} Monte Carlo sampling at $x \approx 1/3$ were a legitimate model of IS geometry relevant to FS hadron production for $\eta \approx 0$ and $x \approx 0.01$.

\begin{figure}[h]
  \includegraphics[width=1.65in,height=1.65in]{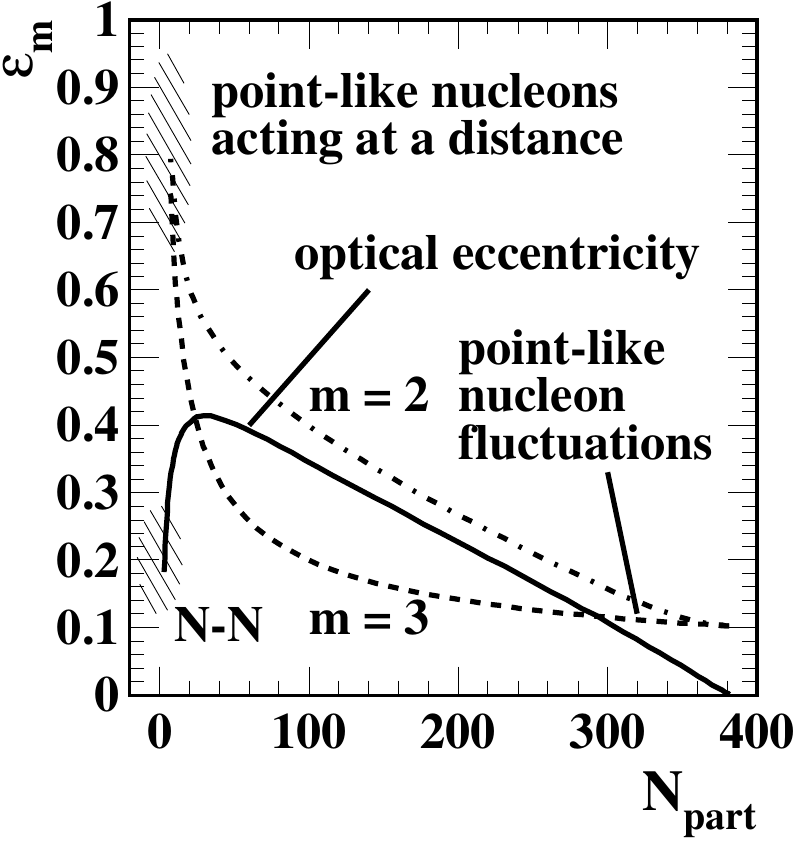}
  \includegraphics[width=1.65in,height=1.65in]{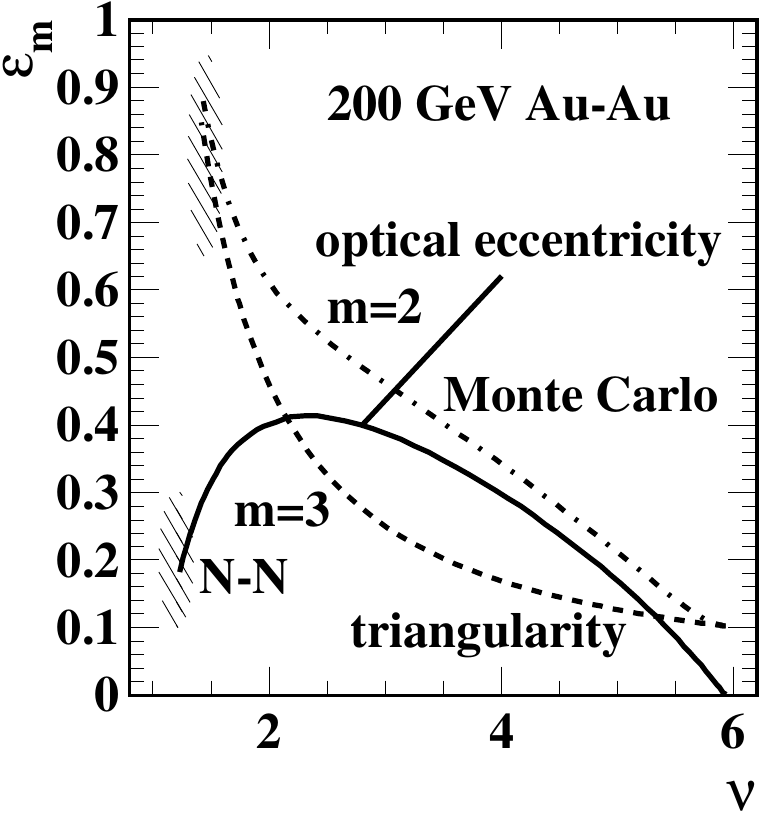}
\caption{\label{epps}
Centrality trends for optical and Monte Carlo Glauber initial-state azimuth multipoles with $m = 2$, 3, on participant-projectile-nucleon number (left panel) and binary \nn\ collisions $\nu$ per participant-nucleon  pair (right panel).
 } 
 \end{figure}

Figure~\ref{epps} shows centrality trends for $m = 2,$ 3 IS power-spectrum elements on participant-nucleon number $N_{part}$ (left panel) and mean participant pathlength $\nu$ (right panel). Plotted are optical eccentricity $\epsilon_{2,opt}$ (solid curves), Monte-Carlo eccentricity $\epsilon_{2,MC}$ (dash-dotted curves) and so-called ``triangularity'' $\delta \epsilon_3$ (dashed curves). 
From Eq.~(\ref{optical}) (and ignoring a possible $\sigma^2_{\epsilon_2 }$contribution) we have $\epsilon^2_{2,MC} = \epsilon^2_{2,opt} + \delta\epsilon^2_2$ with $\delta\epsilon^2_2 \approx 4/N_{part}$ and $\epsilon^2_{3,MC} = \delta\epsilon^2_3 \approx 4/N_{part}$. %

Whether point-like-nucleon sampling at nucleon momentum fraction $x \approx 1/3$ represents an IS geometry with significant manifestations in FS correlation structure for $x \approx 0.01$ is an open question. The present analysis indicates that optical eccentricity $\epsilon_{2,opt}$ is most compatible with NJ $v_2\{\text{2D}\}$ data obtained from 2D model fits.

\section{Nongraphical numerical methods} \label{nongraph}

In this section we present a simplified analysis method confined to 1D azimuth difference $\phi_\Delta$ corresponding to projection  $\rho(\eta_\Delta,\phi_\Delta) \rightarrow \rho(\phi_\Delta)$ of all 2D angular correlation structure within some detector $\eta$ acceptance $\Delta \eta$. This description is directly related to conventional 1D $v_2$ analysis via {\em nongraphical numerical methods} (NGNM). In Sec.~\ref{modelfit} we describe a more general analysis method based on model  fits to 2D data histograms on $(\eta_\Delta,\phi_\Delta)$, the basis for the present NJ-quadrupole study.


\subsection{Pair densities on azimuth difference $\bf \phi_\Delta$}


$\rho(\phi_\Delta)$ is the (4D angular) pair-density projection onto azimuth difference  $\phi_\Delta$ obtained by {\em averaging} over $\eta_\Delta$ within acceptance $\Delta \eta$. Any distribution on periodic $\phi_\Delta$ can be described exactly by a Fourier series
\bea \label{simpleauto}
\rho(\phi_\Delta) \hspace{-.005in} &=& \frac{1}{2\pi \Delta \eta^2} \overline{\sum_{i,j = 1,\in\Delta \eta}^{n} \delta(\phi_i - \phi_j - \phi_\Delta)} \\ \nonumber
&=&\hspace{-.005in} \delta(\phi_\Delta) \rho_0 / \Delta \eta \hspace{-.002in} +\hspace{-.002in} V_0^2 \hspace{-.002in}+\hspace{-.002in} 2 \hspace{-.002in} \sum_{m=1}^\infty \hspace{-.002in} V_m^2 \cos(m\phi_\Delta),
\eea
with $ V_m^2 \equiv  V_0^2v_m^2 $ and $V_0^2 = \overline{n(n-1)}/(2\pi \Delta \eta)^2$ (overline indicates event-ensemble average). The sum upper limit is simplified by $n_{ch} \rightarrow n$. That expression describes all pairs within the angular acceptance, including self pairs $i=j$. 
The Fourier coefficients $V_m^2$ constitute the {\em power spectrum} of the azimuth pair density~\cite{lti}.  
Any and all significant correlation structure {\em projected onto azimuth} should be accurately represented by the $V_m^2$, which are additive or extensive measures (whereas the $v_m^2$ are not). 
The delta function on the RHS represents self pairs ($i = j$) and has its own Fourier series representation (a uniform or ``white-noise'' power spectrum on $m$)~\cite{lti}. 

In this analysis we consider the azimuth quadrupole (a cylindrical multipole) term of the power spectrum.
The $m=2$ Fourier coefficient (also a 4D angular density) with {\em self-pairs excluded}  is determined from data by
\bea \label{bigv2}
V_2^2(b) &=& \frac{1}{2\pi}  \int_0^{2\pi} d\phi_\Delta \, \rho(\phi_\Delta,b) \cos(2 \phi_\Delta) \\ \nonumber
&=& \frac{1}{(2\pi \Delta \eta)^2}  \overline{\sum_{i  \neq j = 1, \in \Delta \eta}^{n,n-1}  \cos[2(\phi_i - \phi_j)] \rangle} \\ \nonumber
&\equiv& \rho_0^2(b) v_2^2(b) \approx \rho_0^2(b) \langle \cos(2\phi_\Delta) \rangle.
\eea
The approximation in the last line is for $n \gg 1$. 
For conventional notation $v_2\{\text{method}\}$ Eq.~(\ref{bigv2}) represents $V_2^2\{2\}$ derived from two-particle 1D azimuth correlations.
Recent ``higher-harmonic'' Fourier analysis extends to $m > 2$, interpreting any $v_m$ as representing a flow~\cite{gunther}. Equation~(\ref{bigv2}) is actually equivalent to a model fit to 2D angular correlations projected onto 1D azimuth. Model-fit comparisons are discussed further in Sec.~\ref{modelfit}.

\subsection{$\bf y_t$-differential $\bf V_2^2\{2\}(y_t,b)$ measurement}

 Figure~\ref{kine} (right panel) shows joint and marginal conditions on $(y_{t1},y_{t2})$ (bold lines). The {\em joint} distribution (6D density) on  $(y_{t1},y_{t2})$ is defined by
\bea
 V_2^2(y_{t1},y_{t2},b) 
 \hspace{-.05in}&=& \hspace{-.05in}\frac{1}{(\delta y_t 2\pi \Delta \eta)^2} \hspace{-.08in} \overline{\sum_{i \in y_{t1} \neq j \in y_{t2}}^{n_{y_{t1}},n_{y_{t2}}} \hspace{-.15in} \frac{\cos[2(\phi_i - \phi_j)]}{y_{t1}y_{t2}}},\,\,
\eea
where $\delta y_t = 0.4$ is the bin width on $y_t$ (uniform widths in this analysis). The {\em marginal} distribution is defined by
\bea \label{v222}
 V_2^2(y_{t},b) &=& \frac{1}{2\pi}  \int_0^{2\pi} d\phi_\Delta \, \rho(y_t,\phi_\Delta,b) \cos(2 \phi_\Delta) \\ \nonumber
&=&\frac{1}{\delta y_t(2\pi \Delta \eta)^2} \overline{\sum_{i \in y_{t}\neq j=1 }^{n_{y_t},n-1} \frac{\cos[2(\phi_i - \phi_j)]}{y_{t}}} \\ \nonumber
&=& \int_0^\infty dy_t' y_t' V_2^2(y_{t},y_{t}',b) \\ \nonumber
&\equiv&V_2(b) V_2(y_{t},b)  \\ \nonumber
&=& \rho_0(b) \rho_0(y_t,b) v_2^2(y_t,b),
\eea
where $\rho_0(y_t,b)$ is the SP spectrum on $y_t$. 
$V_2(y_t,b) \equiv V_2^2(y_t,b)/V_2(b)$ with $V_2(b) \equiv \sqrt{V_2^2(b)}$ defines a self-consistent extensive measure system.

In some conventional NGNM $v_2$ analyses the pair ratio is calculated directly as the ensemble mean of a ratio
\bea \label{v22ytb}
 v_2^2(y_{t},b) 
&=&\overline{\frac{1}{n_{y_t} (n-1)} \sum_{i \in y_{t}\neq j=1 }^{n_{y_t},n-1} \cos[2(\phi_i - \phi_j)]} 
\\ \nonumber
&\equiv&v_2(b) v_2(y_{t},b),
\eea
whereas in other analyses a ratio of mean values is employed. 
The pair ratio does cancel detector imperfections to some extent. Quadrupole amplitudes $V_2^2(y_{t},b)$ can be approximated by using corrected $y_t$ spectra $\rho_0(y_t,b)$ and angular densities $\rho_0(b)$ averaged over $2\pi$ and some $\eta$ acceptance $\Delta \eta$ (e.g.\ STAR TPC). Data from NGNM applied to two-particle correlations are denoted by $v_2\{2\}$. The $v_2\{\text{2D}\}$ method used in the present study and based on model fits to 2D angular correlations is described next.

\section{Model fits to 2D $\bf (\eta,\phi)$ correlations} \label{modelfit}

In this section we extend NGNM analysis of 1D azimuth projections to {\em graphical} analysis of 2D angular autocorrelations via fits to data with a 2D model function including several elements. The $\eta$ dependence of angular correlations is used to distinguish among several  functional forms that are later interpreted physically by comparisons with theory predictions. 
A {\em per-particle} model appropriate for studying Glauber linear superposition in the context of jet production~\cite{anomalous} with amplitudes denoted by quantities $A_X$ was applied in a previous $y_t$-integral $v_2\{\text{2D}\}$ analysis~\cite{davidhq,noelliptic}. A {\em per-pair} model with amplitudes denoted by $B_X$ is applied in the present $y_t$-differential analysis to provide direct comparison with $v_2(p_t)$ data from conventional 1D analysis. The $B_x$ format does not require corrected SP spectra $\rho_0(y_t,b)$. 



\subsection{Motivation for 2D model fits} \label{modelmotives}

NGNM data (e.g., $v_2\{2\}$) derived from 1D projections discard the $\eta$ structure of 2D angular correlations. Although a Fourier series can represent accurately any 1D azimuth distribution, multiple physical mechanisms may then contribute to any single Fourier amplitude (azimuth multipole)~\cite{bayes}.
%
1D Fourier analysis may be unsuited to describe 2D angular correlations from \pp\ and peripheral \aa\ collisions which include strong variations on $\eta_\Delta$ accurately described by a combination of 1D and 2D peaked functions~\cite{porter2,porter3}. In that case a 1D Fourier series cannot describe angular correlations comprehensively over all \aa\ centralities as required to understand the centrality evolution of collision phenomena~\cite{multipoles}.

A 2D data model composed of elementary functions can remove ambiguities arising from 1D projection onto azimuth.
The basic premise is as follows: Within some limited $\eta$ acceptance 2D structure is separated into what is approximately uniform on $\eta$ and what is strongly varying. The nearly-uniform components are {\em candidates} for 1D Fourier representation. However, alternative representations (e.g., 1D Gaussian on azimuth) are also considered. Any components nonuniform on $\eta$ should be modeled by the simplest combination of elementary functions {\em sufficient} to describe the 2D data accurately. 


\subsection{Isolating and modeling 2D correlation structure}

If Eq.~(\ref{simpleauto}) is evaluated for particle pairs drawn from the same event the result is the sibling-pair density $\rho_{sib}$. If pairs are drawn from different but similar events the resulting mixed-pair density $\rho_{mix}$ is approximately equal to the factorized reference density $\rho_{ref} = V^2_{0,ref} = \rho^2_0$.
In that context the density of ``correlated pairs'' (a covariance density {\em with self pairs excluded}) is
\bea \label{nf}
\Delta \rho(\phi_\Delta) \hspace{-.06in} &=& \hspace{-.06in} \rho_{sib} - \rho_{ref} = \Delta V_0^2 +   2\sum_{m=1}^\infty \hspace{-.02in} V_m^2 \cos(m\phi_\Delta), 
\eea
and the per-pair measure of correlated pairs is pair ratio
\bea 
\frac{\Delta \rho}{\rho_{ref}} &=&  \Delta v_0^2 +  2\sum_{m=1}^\infty \hspace{-.002in} v_m^2 \cos(m\phi_\Delta).\eea
Note that $v_1^2$ in that series represents a {\em cylindrical} multipole, not the {\em spherical} multipole associated with ``directed flow'' $v_1$~\cite{2004}. The quantity $\Delta V_0^2 = V_0^2 - V_{0,ref}^2$ is proportional to fluctuation measure $\sigma_n^2 - \bar n$ (variance difference) representing number angular correlations with characteristic lengths comparable to or exceeding the acceptance scale~\cite{inverse,ptscale}.  We wish to extend the mathematical representation of Eq.~(\ref{nf}) to 2D angular correlations. 

In all 2D data histograms we observe an AS structure (AS ridge) that is broad on azimuth and {\em approximately uniform on $\eta_\Delta$} within the TPC angular acceptance. The latter property implies that the AS structure is a candidate for 1D Fourier representation, but the former property implies that only a few terms in the Fourier series  of Eq.~(\ref{nf}) are required by the data, and two terms (dipole and quadrupole) are sufficient in most cases. An $m=1$ AS dipole is generally consistent with pQCD jet structure in minimum-bias angular correlations, and the $m=2$ NJ quadrupole is the object of the present study.

The remaining 2D structure is strongly varying on $\eta_\Delta$ and therefore not suitable for 1D Fourier series representation.  The $\eta$-dependent 2D structure is represented by a {\em non-Fourier} (NF) term. The model function is then
\bea \label{nfeq}
{\Delta \rho}(\eta_\Delta,\phi_\Delta) \hspace{-.02in} &\equiv& \hspace{-.02in} \Delta \rho_\text{NF}(\eta_\Delta,\phi_\Delta) \hspace{-.02in} + \hspace{-.02in} 2 \sum_{m=1}^2 V_m^2 \cos(m \phi_\Delta),~~~~
\eea
where ${\Delta \rho_\text{NF}}$ is a combination of 1D (on $\eta_\Delta$) and 2D peaked functions plus constant offset. 

In \pp\ collisions the NF contribution dominates 2D angular correlations and consists of elements predominantly associated  with either like-sign charge pairs or unlike-sign pairs (whereas the NJ quadrupole is observed to include both types equally)~\cite{porter2,porter3}. NF includes a 1D peak on $\eta_\Delta$ nearly uniform on $\phi_\Delta$ and a complex SS  2D peaked structure at the $(\eta_\Delta,\phi_\Delta)$ origin. Most of the SS 2D peak can be modeled by a 2D Gaussian consistent in its properties with pQCD jet expectations~\cite{jetsyield}. A smaller 2D {\em exponential} contribution is consistent with Bose-Einstein correlations (BEC) plus conversion-electron pairs~\cite{anomalous}. 

In more-central \auau\ collisions the minimum-bias SS 2D peak becomes elongated on $\eta_\Delta$ and slightly narrower on $\phi_\Delta$ compared to \pp\ collisions but remains statistically consistent with a 2D Gaussian~\cite{anomalous}. A NJ quadrupole is visually obvious in more-central \auau\ data~\cite{axialci,anomalous}, but the quadrupole  component remains statistically significant for all \auau\ centralities down to \nn\ collisions~\cite{davidhq}. 
Thus, parametric evolution of a {\em single} 2D model function with a few simple elements accurately represents all collision systems from \pp\ to central \auau. This study presents a description of model properties inferred from 2D histograms by $y_t$-differential model fits.

\subsection{$\bf y_t$-differential model function}

The \yt-integral $v_2\{\text{2D}\}(b)$ analysis described in Refs.~\cite{davidhq,noelliptic} employed an 11-parameter model function as described in Ref.~\cite{anomalous}. For the present $y_t$-differential analysis we introduce three changes to the fit model: (a) We switch from per-particle measure $\Delta \rho / \sqrt{\rho_{ref}}$ to per-pair measure $\Delta \rho / \rho_{ref}$ to facilitate comparisons with published $v_2(p_t)$ data from 1D methods while not requiring corrected SP spectra $\rho_0(y_t,b)$. 
(b) We eliminate one NF model element not required to describe more-central \auau\ collisions.  (c) We introduce a Gaussian alternative model for the AS 1D peak.

With $y_t$ cuts imposed jet-related peak structures may become narrower, as expected for jet correlations and observed in so-called trigger-associated dihadron correlations~\cite{trigger,starridge}. If the AS 1D peak on azimuth narrows its Fourier series representation may require more than a single AS dipole term, and an AS 1D Gaussian may be a more efficient representation. A narrower AS 1D peak may also introduce a systematic ambiguity between the NJ quadrupole and a jet-related quadrupole contribution from the narrower AS peak~\cite{tzyam}. In this analysis each of two AS 1D peak models is included alternately in 2D model fits. Any differences in inferred $B_Q\{\text{2D}\}(y_t,b)$ values provide an estimate of systematic uncertainties.

Given that context we simplify the 2D data model for several reasons: (a) The $y_t$-differential analysis includes 99 histograms each for 62 and 200 GeV and for two AS peak models. Each of nearly 400 conditions then requires up to 1000 fits with random starting parameters to insure location of {\em global} $\chi^2$ minima. The entire analysis program is executed several times to investigate data quality, alternative data models and overall fit stability. Of order one million fits are then required. (b) With data subdivided into 11 centrality bins and 9 $y_t$ bins the statistical power for each fit is reduced. (c) In this analysis we emphasize quadrupole systematics for more-central \auau\ collisions where some of the NF terms in Eq.~(\ref{nfeq}) are not required for accurate data description. For those reasons a model with fewer parameters is permitted and provides improved fit stability and more rapid convergence. 

The per-particle model in Ref.~\cite{anomalous} includes eleven parameters, of which only five or six parameters represent physically relevant model elements [SS 2D peak (3), AS 1D peak (1 or 2), NJ quadrupole (1)]. The remainder are mainly required for structure prominent only in peripheral \aa\ and \pp\ collisions. 
The soft-component term $A_{soft} \exp\{-\eta_\Delta^2 / 2 \sigma^2_s\}$ (two parameters), included in the present study for more-peripheral collisions, is not required for more-central \auau\ collisions. In $y_t$-integral analysis the amplitude of that term is observed to fall to zero above mid centrality (50\% fractional cross section)~\cite{anomalous}. 
We drop the model element representing BEC + electron pairs (three parameters) because that structure becomes very narrow in more-central \auau\ collisions.  To compensate, three histogram bins near the ($\eta_\Delta,\phi_\Delta$) origin are removed from the 2D fits to eliminate sensitivity to the narrow BEC + electron-pair component.

The 2D model function for $y_t$-differential analysis applicable to more-central \aa\ collisions is then~\cite{davidhq2}
\bea \label{estructfit}
\frac{\Delta \rho}{\rho_{ref}} \hspace{-.02in}  & = &  \hspace{-.02in}
B_0+  B_{2D} \, \exp \left\{- \frac{1}{2} \left[ \left( \frac{\phi_{\Delta}}{ \sigma_{\phi_{\Delta}}} \right)^2 \hspace{-.05in}  + \left( \frac{\eta_{\Delta}}{ \sigma_{\eta_{\Delta}}} \right)^2 \right] \right\} \nonumber \\
&+&  \hspace{-.02in} B_{D}\, \{1 +\cos(\phi_\Delta - \pi)\} / 2 + \hspace{-.02in}  B_{Q}\, 2\cos(2\, \phi_\Delta),
\eea
where $B_0$ is a constant offset and $B_Q \rightarrow B_Q\{\text{2D}\} = v_2^2\{\text{2D}\}$ denotes the NJ quadrupole derived from model fits to 2D angular correlations.
The soft-component term with amplitude $B_\text{soft}$ does not appear explicitly in Eq.~(\ref{estructfit}) but is included in model fits to more-peripheral data histograms. Equation~(\ref{estructfit}) is then referred to as the {\em eight-parameter model}.

As noted, the AS 1D peak can be modeled by an AS 1D Gaussian or by its limiting case, an AS dipole [as in Eq.~(\ref{estructfit})]~\cite{tzyam,multipoles}. If the AS dipole is chosen the inferred quadrupole amplitude determines an upper limit on the true {\em nonjet} quadrupole. If the AS 1D Gaussian is chosen the inferred quadrupole determines a lower limit. The difference estimates systematic uncertainties for $v_2^2\{\text{2D}\}(y_t,b)$.  The 2D fit model also includes image Gaussians at $2\pi$ for the SS 2D Gaussian and at $-\pi$ for the AS 1D Gaussian if that peak model is employed~\cite{tzyam}.

Amplitudes are denoted by symbol $B_X$ in the 2D fit model of Eq.~(\ref{estructfit}) applied to per-pair data histograms $\Delta \rho / \rho_{ref}$ as opposed to amplitudes  $A_X$ for per-particle histograms. From model fits to 2D angular correlations we measure pair ratio $
B_Q\{\text{2D}\}(y_t,b) = v_2^2\{\text{2D}\}(y_t,b)$ and may convert those data to power-spectrum elements $V_2^2(y_t,b)$ via corrected SP $y_t$ spectra 
$\rho_0(y_t,b)$ (3D densities) 
and yields $\rho_0(b)$ (2D densities). 


 \section{2D angular autocorrelations} \label{2dang}

In this section we introduce data histograms as 2D angular autocorrelations. The basic analysis procedure is described in Ref.~\cite{anomalous}. 
Charged hadrons from \auau\ collisions at $\sqrt{s_{NN}} = 62$ and 200 GeV accepted for this analysis fell within a detector acceptance defined by $p_t>0.15$~GeV/$c$, $|\eta| < 1.0$ and $2 \pi$ azimuth. 
Charge signs were determined but particle identification was not otherwise implemented.  Further details of track definitions, efficiencies and quality cuts are described in Ref.~\cite{anomalous}.  
Data for this analysis were selected from earlier RHIC running periods where low luminosities insure reduced pileup distortions in 2D angular correlations. Reduction of pileup effects and other tracking details are described in Ref.~\cite{anomalous}. In this presentation we emphasize the 200 GeV data; the results for 62 GeV are similar modulo the energy-dependence factor reported in Refs.~\cite{davidhq,noelliptic}.

\subsection{$\bf y_t$-differential 2D histograms}

\begin{figure}[t]
 \includegraphics[width=1.65in,height=1.58in]{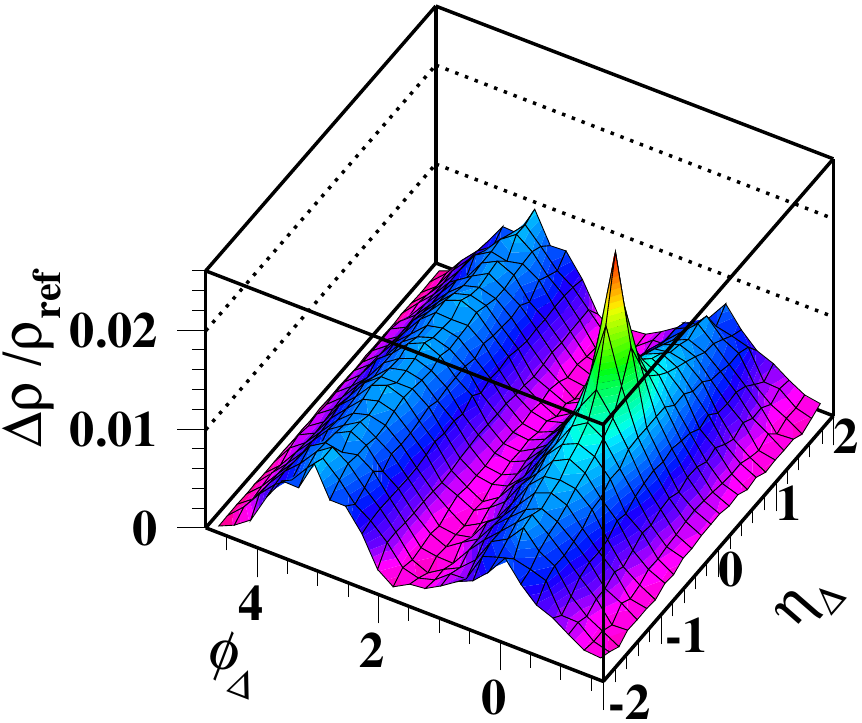}
\put(-100,95) {\bf (a)}
\includegraphics[width=1.65in,height=1.58in]{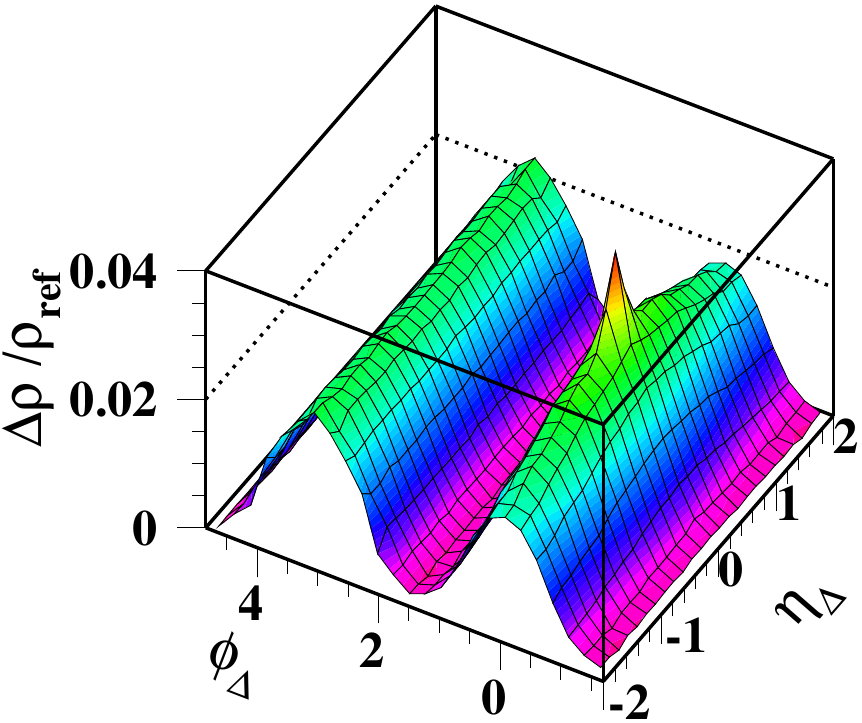}
\put(-140,105) {\bf 200 GeV}
\put(-100,95) {\bf (b)} \\
 \includegraphics[width=1.65in,height=1.58in]{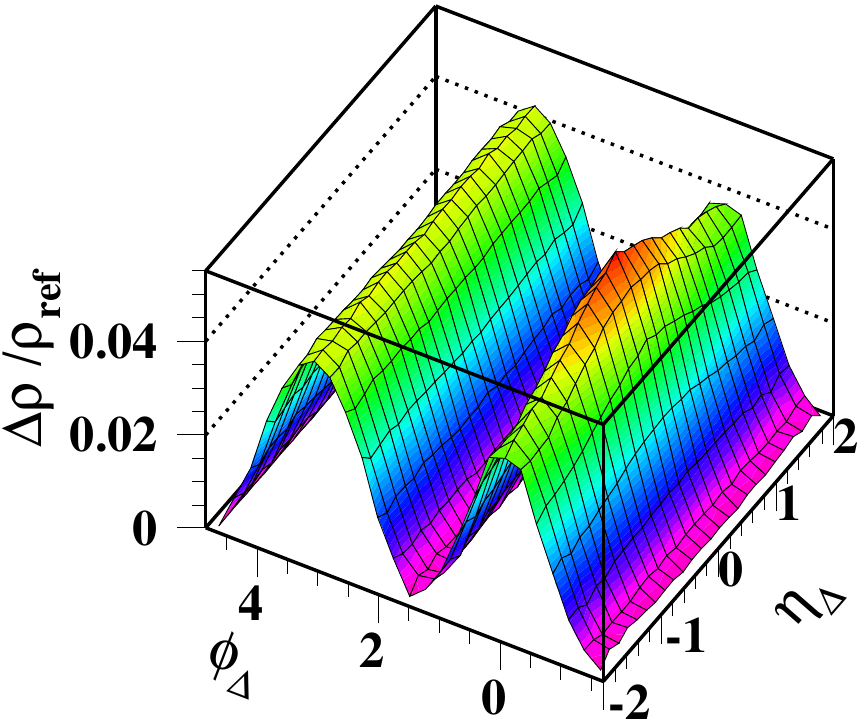}
\put(-100,95) {\bf (c)}
\includegraphics[width=1.65in,height=1.58in]{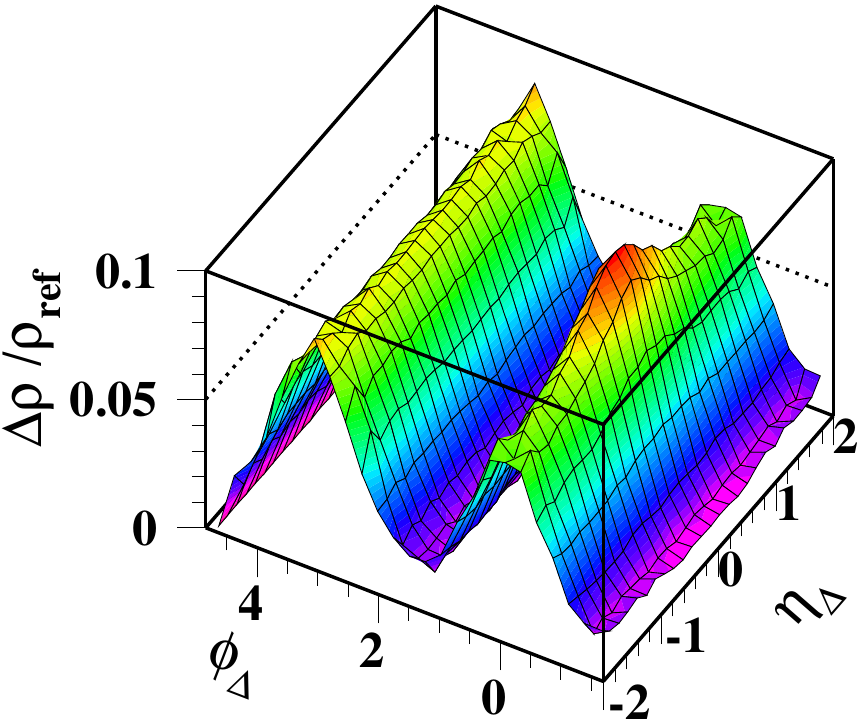}
\put(-100,95) {\bf (d)}
 \caption{\label{ptdiff1} (Color online)
2D data histograms from 40-50\% central 200 GeV \auau\ collisions for four $y_t$ bins: (a) $y_t < 1.4$, (b) $1.8 < y_t < 2.2$, (c) $2.6 < y_t < 3.0$, (d) $3.4 < y_t < 3.8$.  The narrow BEC + electron peak at the origin decreases to zero amplitude for larger $y_t$.
}
\end{figure}

Figure~\ref{ptdiff1} shows example $y_t$-differential  2D angular autocorrelations for nominal 40-50\%-central 200 GeV \auau\ collisions. The mid-central case is chosen to illustrate typical model fits over a range of $y_t$ bins. The $y_t$-differential analysis is based on 6.7M and 14.5M (both year 2004) Au-Au collisions at $\sqrt{s_{NN}} =$ 62 and 200 GeV respectively. The panels represent \yt\ intervals (a) $y_t < 1.4$, (b) $1.8 < y_t < 2.2$, (c) $2.6 < y_t < 3.0$ and (d) $3.4 < y_t < 3.8$ (\yt\ bins 1, 3, 5, 7).

The BEC-electron peak dominates correlations in the lowest $y_t$ bin as expected. That peak is wide enough in the lowest bin that it interferes with the SS 2D jet-like peak and degrades the fit quality. Since the information obtained on jet structure in the lowest $y_t$ bin is minimal, fit values for SS 2D peak parameters from that bin are omitted from the rest of the analysis. For larger-$y_t$ bins the BEC-electron peak is reduced in amplitude and widths.  A few histogram bins nearest the origin are removed from the fits and the 2D fit model includes no corresponding element. Nonjet quadrupole data are insensitive to that issue and are  retained for all $y_t$ bins. 

\subsection{Example 2D model fits to $\bf y_t$-differential data}

\begin{figure}[h]
 \includegraphics[width=3.3in,height=1.58in]{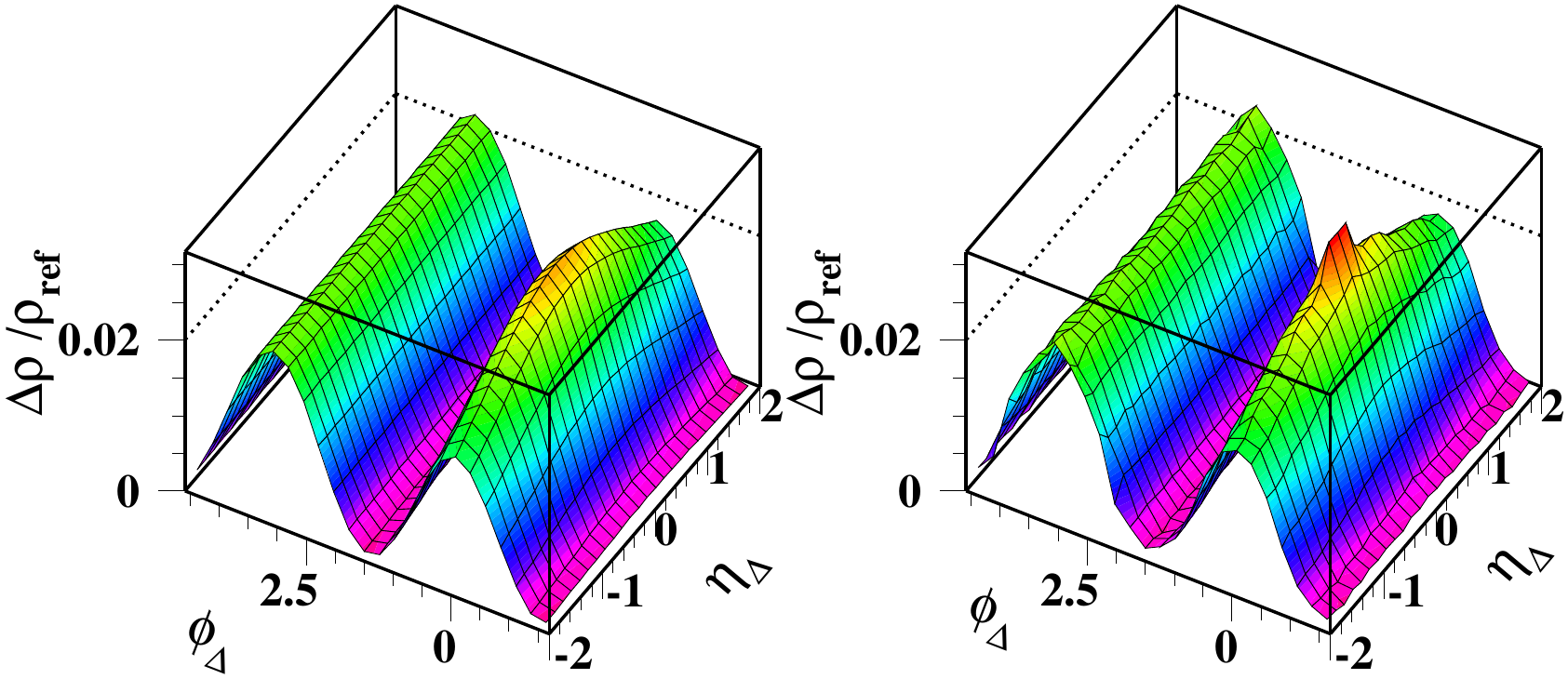}
\put(-100,95) {\bf (b)}
\put(-220,95) {\bf (a)}\\
\includegraphics[width=3.3in,height=1.58in]{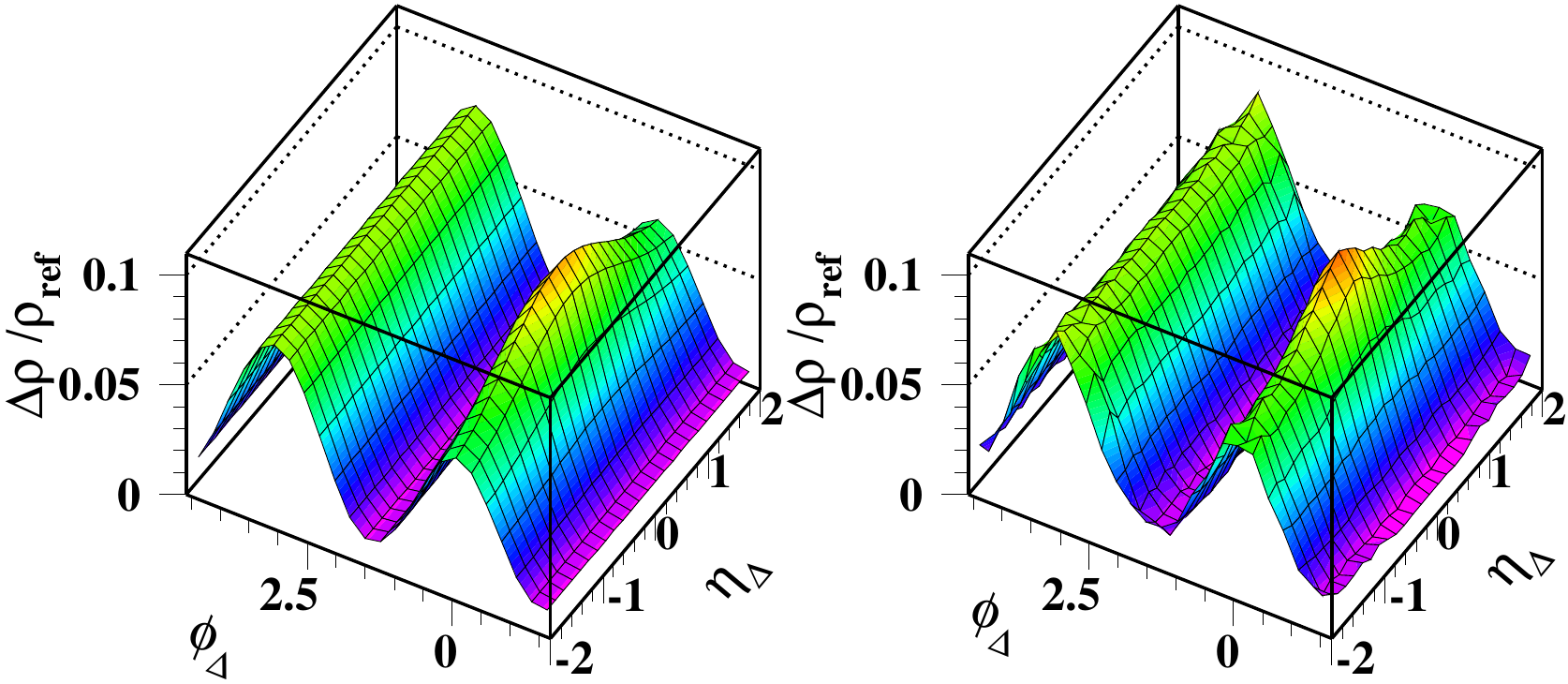}
\put(-100,95) {\bf (d)} 
\put(-220,95) {\bf (c)} 
 \caption{\label{fits} (Color online)
Fits to 2D data histograms from 40-50\% central 200 GeV \auau\ collisions for two $y_t$ bins: (a), (b) $1.8 < y_t < 2.2$ and (c), (d) $3.4 < y_t < 3.8$.  The model fits appear on the left in each case. The vertical scale is the same for fits and data. The upper limit has been adjusted to reveal the BEC + electron contribution excluded from the fit (narrow peak in three central bins).
} 
 \end{figure}

Figure~\ref{fits} compares fits (left panels) with data histograms (right panels) for two $y_t$ bins from 40-50\% central 200 GeV \auau\ collisions. That figure presents two of 99 cases (11 centralities $\times$ 9 $y_t$ bins). For each case approximately 1000 fits starting with randomly-chosen initial parameter values  are performed. The fit corresponding to the global-minimum $\chi^2$ value is then chosen. The fits were performed with the eight-parameter model function of Eq.~(\ref{estructfit}) including an AS dipole 1D peak model or with a nine-parameter model including an AS Gaussian. The three bins at the origin containing the BEC + electron peak were excluded from all fits. That contribution is mainly confined to the lowest \yt\ bin.

Small irregularities appearing near the pair-acceptance boundary $|\eta_\Delta| = 2$ have little effect on 2D fits because the statistical uncertainties are largest there. The SS 2D peak may deviate from an ideal 2D Gaussian for some applied $p_t$ cuts, but for this analysis of the marginal distribution on $y_t$ the deviations are not substantial. The NJ-quadrupole component is found to be insensitive to such deviations (see Sec.~\ref{uncert} for further discussion).

 \section{$\bf y_t$-differential $\bf B_X\{\text{2D}\}$ Systematics} \label{ytdiffsyst}

\begin{figure}[t]
  \includegraphics[width=1.65in,height=1.62in]{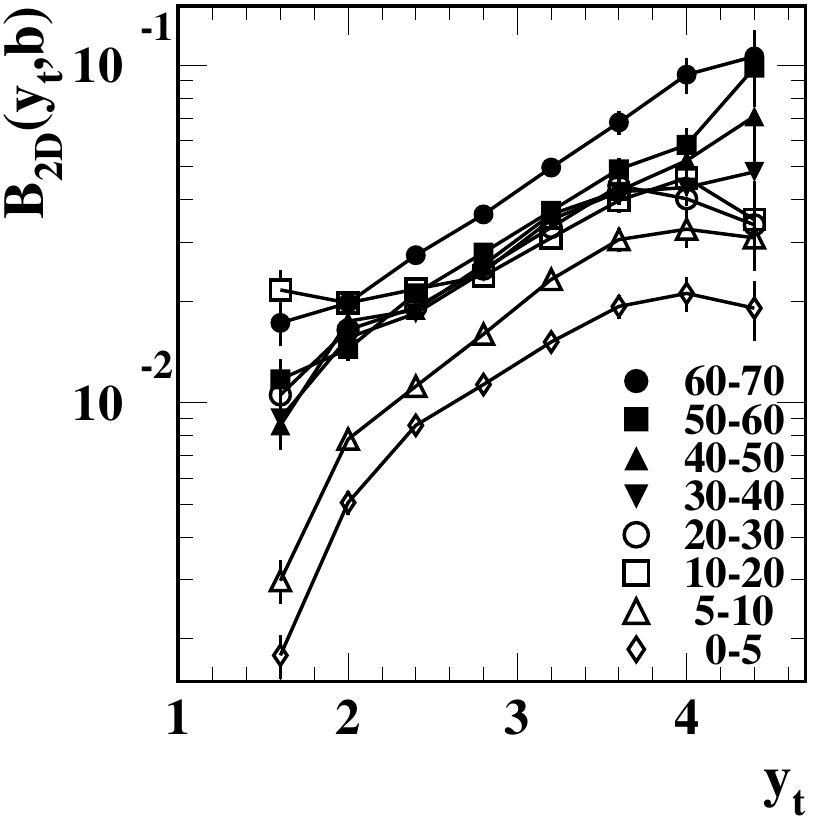}
 \put(-50,99) {\bf (a)}
\includegraphics[width=1.65in,height=1.65in]{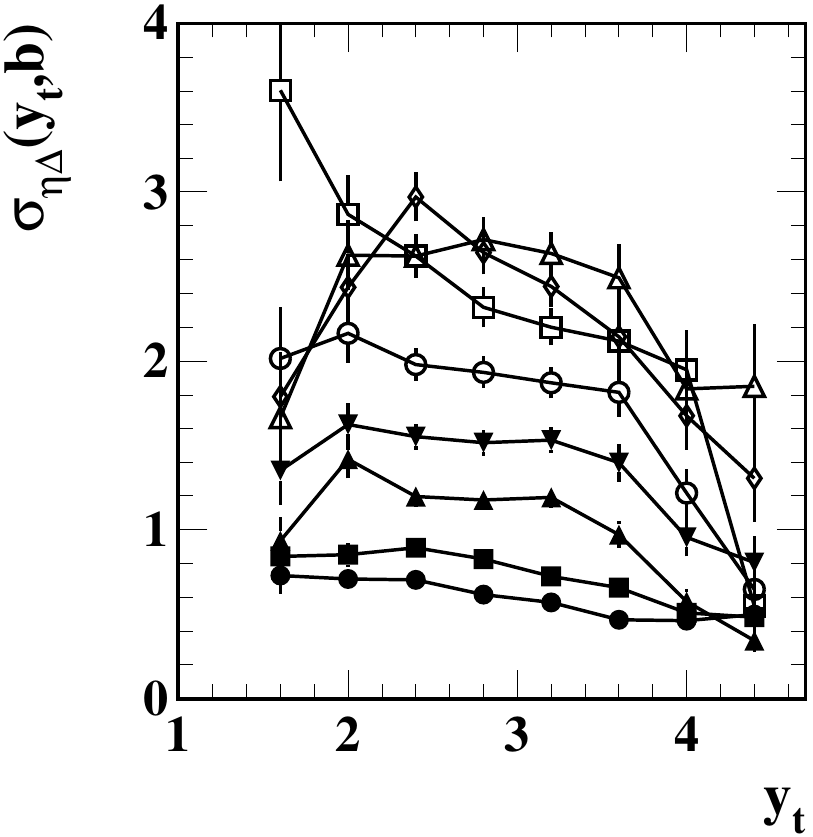}
 \put(-50,99) {\bf (b)} \\
 \includegraphics[width=1.65in,height=1.65in]{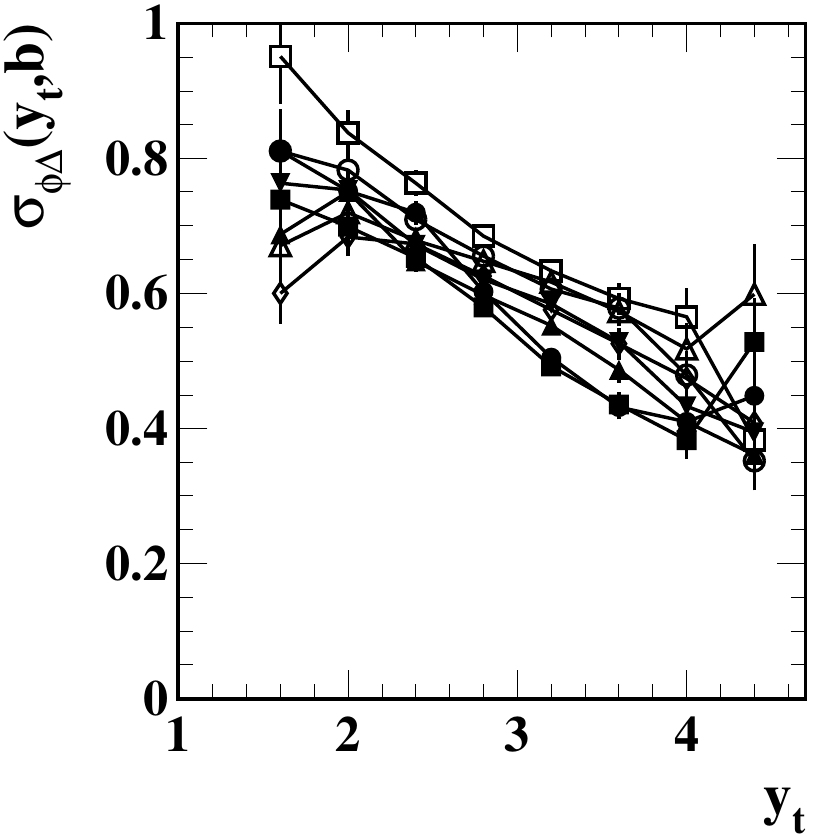}
 \put(-50,99) {\bf (c)}
  \includegraphics[width=1.65in,height=1.65in]{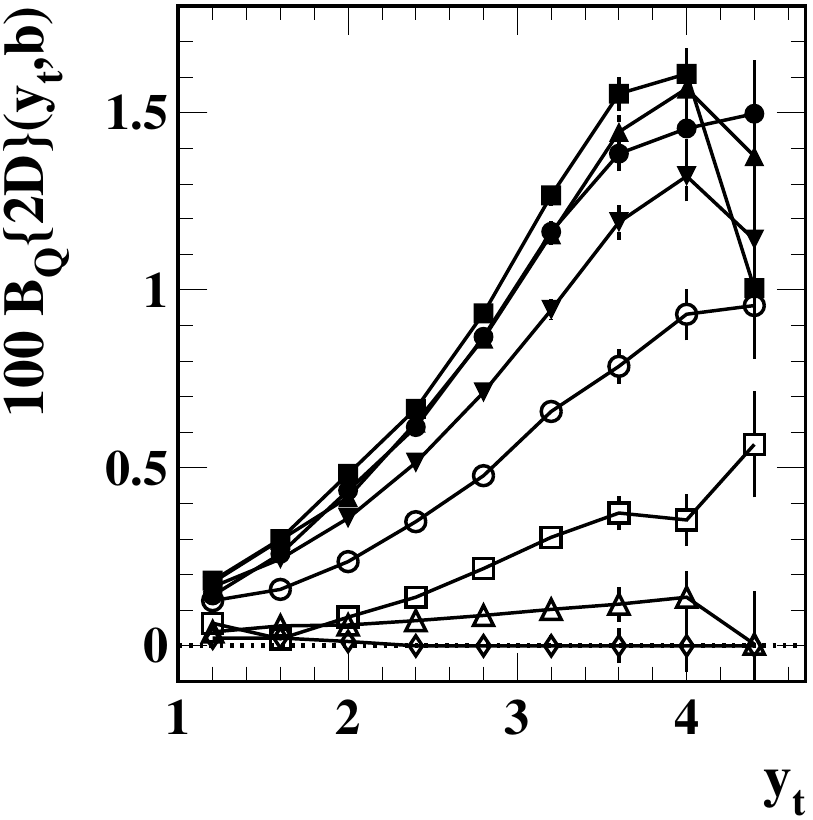}
 \put(-50,99) {\bf (d)}
\caption{\label{data1}
Per-pair fit parameters derived from 2D fits with Eq.~(\ref{estructfit}) to $y_t$-differential data histograms from 200 GeV \auau\ collisions for eight centrality bins (0-70\% central). 
(a) SS 2D peak amplitude $B_{2D}(y_t,b)$, 
(b, c) SS 2D peak r.m.s.\ widths $\sigma_{\eta_\Delta}(y_t,b)$ and $\sigma_{\phi_\Delta}(y_t,b)$,
(d) Per-pair quadrupole amplitude $B_Q\{\text{2D}\}(y_t,b)$. 
Error bars indicate fit uncertainties. 
 } 
\end{figure}

Figure~\ref{data1} shows the $y_t$ and centrality evolution of four fit parameters from 200 GeV \auau\ collisions describing SS 2D peak properties and the NJ quadrupole amplitude. The plots show per-pair SS 2D peak amplitude $B_{2D}$, SS peak widths $\sigma_{\eta_\Delta}$ and $\sigma_{\phi_\Delta}$ and per-pair NJ quadrupole amplitude
$B_Q\{\text{2D}\}$. 
The data for SS 2D peak properties are not plotted for the first $y_t$ bin ($p_t \approx 0.2$ GeV/c) because correlation structure for that centrality is dominated by the narrow peak representing BEC and conversion electrons [see Fig.~\ref{ptdiff1} (a)].  The $y_t$ interval for this analysis includes almost all jet correlation structure. The $y_t$-differential SS 2D peak parameters are required for the $v_2\{2\}(y_t,b) \leftrightarrow v_2\{\text{2D}\}(y_t,b)$ comparisons in Sec.~\ref{ytcomp}.  These results are consistent with Ref.~\cite{davidhq2}.

 The $B_{2D}$ and $B_Q$ data are dominated by trends generally expected for per-pair amplitudes: (a) decrease with increasing centrality and (b) increase with increasing $y_t$, both due to the SP $\rho_0(y_t,b)$ spectrum factor in the denominators of the $B_X$. Smaller physically-meaningful variations are overshadowed by the dominant per-pair trends. Just as for $A_Q\{\text{2D}\}(b)$ the $B_Q\{\text{2D}\}(y_t,b)$ data exhibit remarkable simplicity, but the simplicity is not revealed until we present these results in alternative plotting contexts. 
Note that $B_Q\{\text{2D}\}$ values for 0-5\% central collisions are consistent with zero for almost all $y_t$ bins (at both collision energies), with small upper limits.

In Refs.~\cite{davidhq,noelliptic} it was reported that $y_t$-integral per-particle NJ quadrupole amplitude  $A_Q\{\text{2D}\}(b,\sqrt{s_{NN}})$ is factorizable, the factors having simple functional forms. With the $y_t$-differential data from this study we demonstrate that the $y_t$ or $p_t$ dependence of the NJ quadrupole is also factorizable, leading to a simple quadrupole parametrization accurate over a large kinematic space as presented in Sec.~\ref{v2ytparam}.

The sharp transition in jet-related angular structure near 50\% \auau\ centrality ($\nu \approx 3$)~\cite{anomalous} does not significantly alter the $p_t$ structure of the SS 2D peak. Similar $y_t$ trends are observed down to \pp\ collisions for SS 2D peak and quadrupole despite a substantial change in the SS 2D peak angular shape.
To further explore the NJ quadrupole data systematics we introduce the concepts of quadrupole source boost and quadrupole spectrum

\section{Quadrupole source boost} \label{quadboostt}

The broad source-boost distribution reflecting Hubble expansion of a bulk medium assumed in conventional hydro descriptions is contrasted with a narrow quadrupole source-boost distribution inferred from $v_2(y_t)$ data.

\subsection{Theoretical context}

According to hydro descriptions  elliptic flow is an azimuthal modulation of radial flow corresponding to the IS matter eccentricity of non-central \aa\ collisions, the transverse flows arising from large pressure gradients in a dense medium~\cite{hydro2,radial}. Within the hydro narrative most hadrons (especially below 2-3 GeV/c) emerge by ``freezeout'' from the monolithic flowing medium, and each hadron is therefore associated with a particular medium speed or relativistic boost corresponding to its freezeout space-time position. Almost all final-state hadrons should then reflect the same hadron-source boost distribution.

As noted in Ref.~\cite{hydro2} if all hadrons emerged from a cylindrical shell with fixed radial speed (as a limiting case) the SP  \pt\ spectrum should exhibit a peak at nonzero \pt\ and a minimum at zero, reflecting the radial boost of the hadron source. On transverse rapidity \yt\ the spectrum alteration would be especially simple: the hadron spectrum in the stationary lab frame would be the spectrum in the moving boost frame shifted to larger \yt. For $v_2(p_t)$ the consequence would be negative values for \pt\ near zero~\cite{quadspec}. But such trends are not expected due to ``...a more realistic [hadron source] velocity profile, [wherein] the peak in transverse-momentum distribution disappears.'' The ``more-realistic'' velocity profile is approximated by that expected for transverse Hubble expansion of  a flowing bulk medium, a broad distribution on radial speed $\beta_t$ extending from zero to some maximum value. But so-called ``mass ordering'' of $v_2(p_t)$ at lower \pt\ should survive as a manifestation of radial flow.

In Ref.~\cite{quadspec} it was pointed out that the ratio $v_2(p_t)/ p_t$ for several hadron species plotted vs \yt\ with the proper mass for each hadron species reveals a common source boost distribution for identified hadrons from a minimum-bias distribution of \auau\ collisions (centrality-averaged result). The factor $1/p_t$ emerges from a Taylor expansion of the Cooper-Frye expression~\cite{cooper} for the thermal spectrum from a boosted source. Here we consider the centrality dependence of source boosts inferred from $v_2\{\text{2D}\}(y_t)$ data for unidentified hadrons from 62 and 200 GeV \auau\ collisions.

 \subsection{Quadrupole source-boost centrality dependence}

\begin{figure}[h]
  \includegraphics[width=1.65in,height=1.65in]{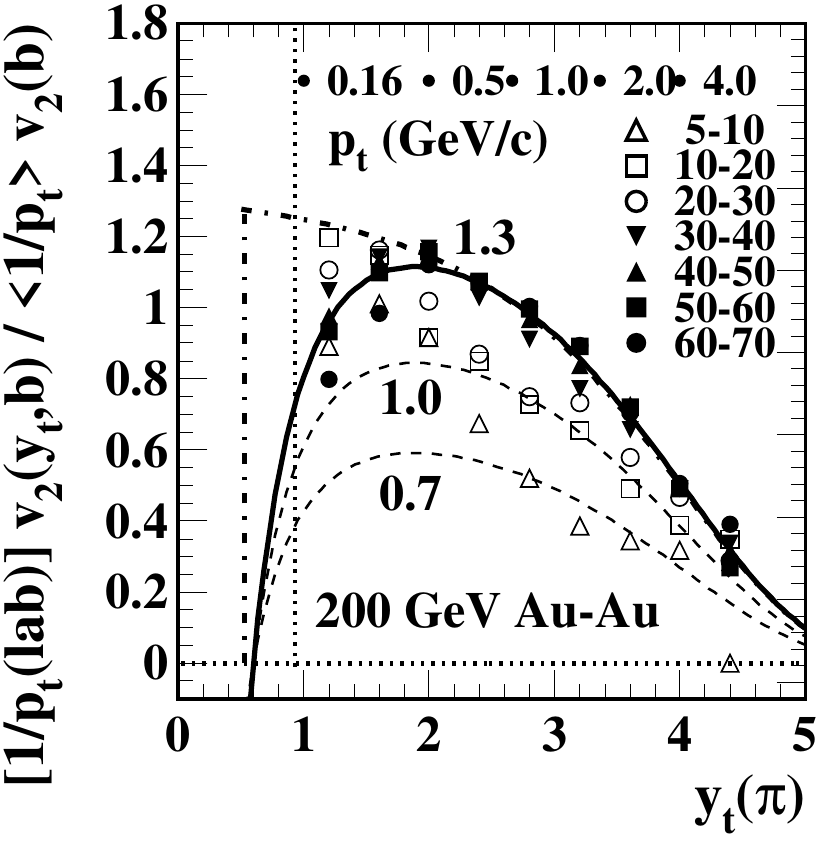}
  \includegraphics[width=1.65in,height=1.65in]{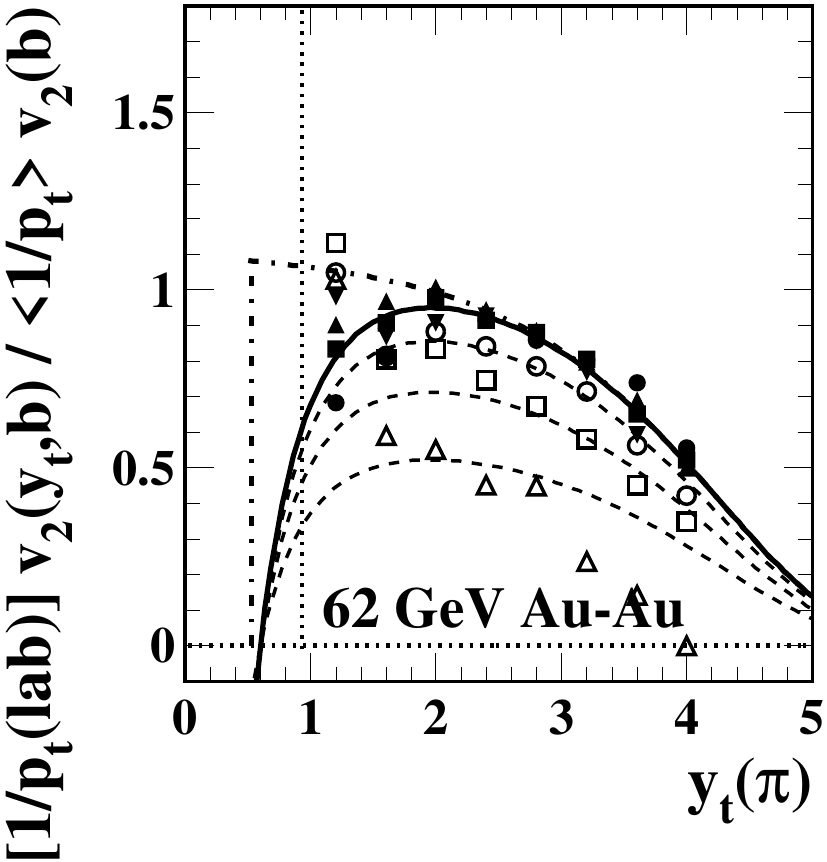}
 \caption{\label{boost2} 
Left:   Per-pair data $B_2\{\text{2D}\}(y_t,b)$ from the present analysis scaled as described in the text and compared with the minimum-bias pion data (solid curve) from Ref.~\cite{quadspec} (left panel). The vertical dotted lines mark the TPC acceptance edge on $y_t$ for pions at $p_t = 0.15$ GeV/c.
Right:  The same procedure applied to data from 62 GeV \auau\ collisions.
} 
 \end{figure}

Figure~\ref{boost2} (left panel) shows the source-boost centrality evolution of $y_t$-differential unidentified-hadron  $B_Q\{\text{2D}\}$ data from Fig.~\ref{data1} (d) for 200 GeV \auau\ collisions. 
The plotted quantity is unit-normal ratio $(1/p_t) B_Q\{\text{2D}\}(y_t,b) / \langle 1/p_t \rangle B_Q\{\text{2D}\}(b)$. The ratio format removes the $y_t$-integral $v_2(b)$ centrality dependence reported in Refs.~\cite{davidhq,noelliptic} [see Eq.~(\ref{v22ytb})] bringing data trends near the spectrum mean $\bar p_t$ ($\bar y_t \approx 1.8$) into alignment at unity. The solid curve represents a centrality-averaged pion curve from Ref.~\ref{quadspec} for quadrupole source boost $\Delta y_{t0} = 0.6$ divided by $\langle 1/p_t \rangle \overline{v_2\{\text{2D}\}(b)} \approx 0.1$ that describes most of the scaled $B_Q\{\text{2D}\}(y_t,b)$ data well. 

Figure~\ref{boost2} (right panel) shows the same procedure applied to  data from 62 GeV \auau\ collisions. The results are very similar, with some small quantitative differences noted below. 
The \yt-integral values $v_2\{\text{2D}\}(b)$ used for both plots are derived from Refs.~\cite{davidhq,noelliptic} for the two energies, and see Sec.~\ref{ytintsyst}. In both cases $\langle 1/p_t \rangle = 2$ (GeV/c)$^{-1}$ and the quadrupole source boost is consistent with $\Delta y_{t0} = 0.6$. 

The solid and dashed curves in Fig.~\ref{boost2} are defined by 
$F(y_t) = g(b) C \{[1-\beta_t / \tanh(y_t)] / (1-\beta_t)\}\exp(-p_t / P)$.
The expression in curly brackets is determined only by relativistic kinematics and $\Delta y_{t0}$~\cite{quadspec}. The other factors are derived from data. 
Factor C is 1.3 and 1.1 respectively for the two energies. Factor $g(b)$ is defined in Sec.~\ref{v2ytparam}.  The product $g(b)C$ for three $g(b)$ values 
(0.55, 0.75, 1.0)  is noted next to the curves (left panel). 
Exponential constant $P$ is $4 \pm 0.2$ GeV/c for 200 GeV and $5 \pm 0.2$ GeV/c for 62 GeV. The increase in $P$ for the lower energy may result from the softer SP spectrum in the $v_2$ denominator: the spectrum hard component at 62 GeV is 60\% of the hard component at 200 GeV~\cite{anomalous,jetsyield}, tending to elevate the plotted $v_2$ ratio at larger $y_t$.

The plotting format in Fig.~\ref{boost2} includes prefactor $1/p_t(\text{lab})$ derived from the measured lab $p_t$. Motivation for that factor relates to interest in the $y_t$ spectrum in the boost frame. In the function $F(y_t)$ the kinematic factor in curly brackets represents the ratio $p_t(\text{boost}) / p_t(\text{lab})$ relating $p_t$ in the lab frame to $p_t$ in the boost frame as derived in Ref.~\cite{quadspec}, Eq.~(15). If the kinematic factor were removed [i.e.\  if $1/p_t(\text{boost})$ were applied as the prefactor] most of the data would follow the dash-dotted curves. 


Within  the conventional hydro narrative one should expect increasing source boosts in more-central \aa\ collisions as IS particle and energy densities, and therefore density gradients, increase. The data in Fig.~\ref{boost2} suggest that all scaled $B_Q(y_t,b)$ data below 20\%-central \auau\ centrality are statistically identical in shape. The ratio data  follow a simple exponential form that facilitates the universal parametrization described in Sec.~\ref{v2ytparam}. Uniformity across most centralities suggests that the quadrupole source boost is approximately independent of \auau\ centrality. Those conclusions are consistent with more-recent Lambda $v_2(p_t)$ data~\cite{newlam} for central \auau\ collisions that show the same source boost as the centrality-averaged data~\cite{v2pions,v2strange}. We pursue that possibility with differential study of quadrupole spectra.

\section{Quadrupole spectra} \label{quadspecc}


We next consider the centrality dependence of azimuth quadrupole spectrum shapes above $p_t = 0.5$ GeV/c. Ratio measure $v_2\{\text{method}\}(y_t,b)$ includes the SP $y_t$ spectrum $\rho_0(y_t,b)$ in its denominator. The SP spectrum has a strong jet contribution (spectrum hard component)~\cite{hardspec} that should not relate to hydro models and is then generally extraneous to the azimuth quadrupole problem. Depending on the $v_2$ method the numerator of $v_2(y_t,b)$ may also include significant contributions from jets in the form of a ``nonflow'' bias. To study the quadrupole spectrum in isolation we remove the jet contributions from numerator and denominator of $v_2$ by focusing on NJ quadrupole amplitude $V_2^2\{\text{2D}\}(y_t,b) = \rho_0(b) \rho_0(y_t,b) B_Q\{\text{2D}\}(y_t,b)$.

Based on $B_Q(y_t,b)$ data described in the previous subsection and Ref.~\cite{quadspec} 
we define a {\em unit-normal} ratio
\bea \label{qb} 
Q(y_t,b) &\equiv &\frac{(1/p_t) V_2^2\{\text{2D}\}(y_t,b)}{\langle 1/p_t \rangle V_2^2\{\text{2D}\}(b)} \\ \nonumber
&\rightarrow& \frac{(1/p_t) \rho_0(y_t,b) v_2\{\text{2D}\}(y_t,b)}{\langle 1/p_t \rangle \rho_0(b) v_2\{\text{2D}\}(b) \times g(b)},
\eea
where 
 the data from the present analysis are of the form $B_Q\{\text{2D}\}(y_t,b) = v_2\{\text{2D}\}(b) v_2\{\text{2D}\}(y_t,b)$. 
We seek the centrality dependence of the quadrupole {\em spectrum shape} represented by $Q(y_t,b)$.
The ad hoc $O(1)$ factor $g(b)$ in the second line is defined and discussed in Sec.~\ref{v2ytparam}.

\begin{figure}[h]
  \includegraphics[width=3.3in]{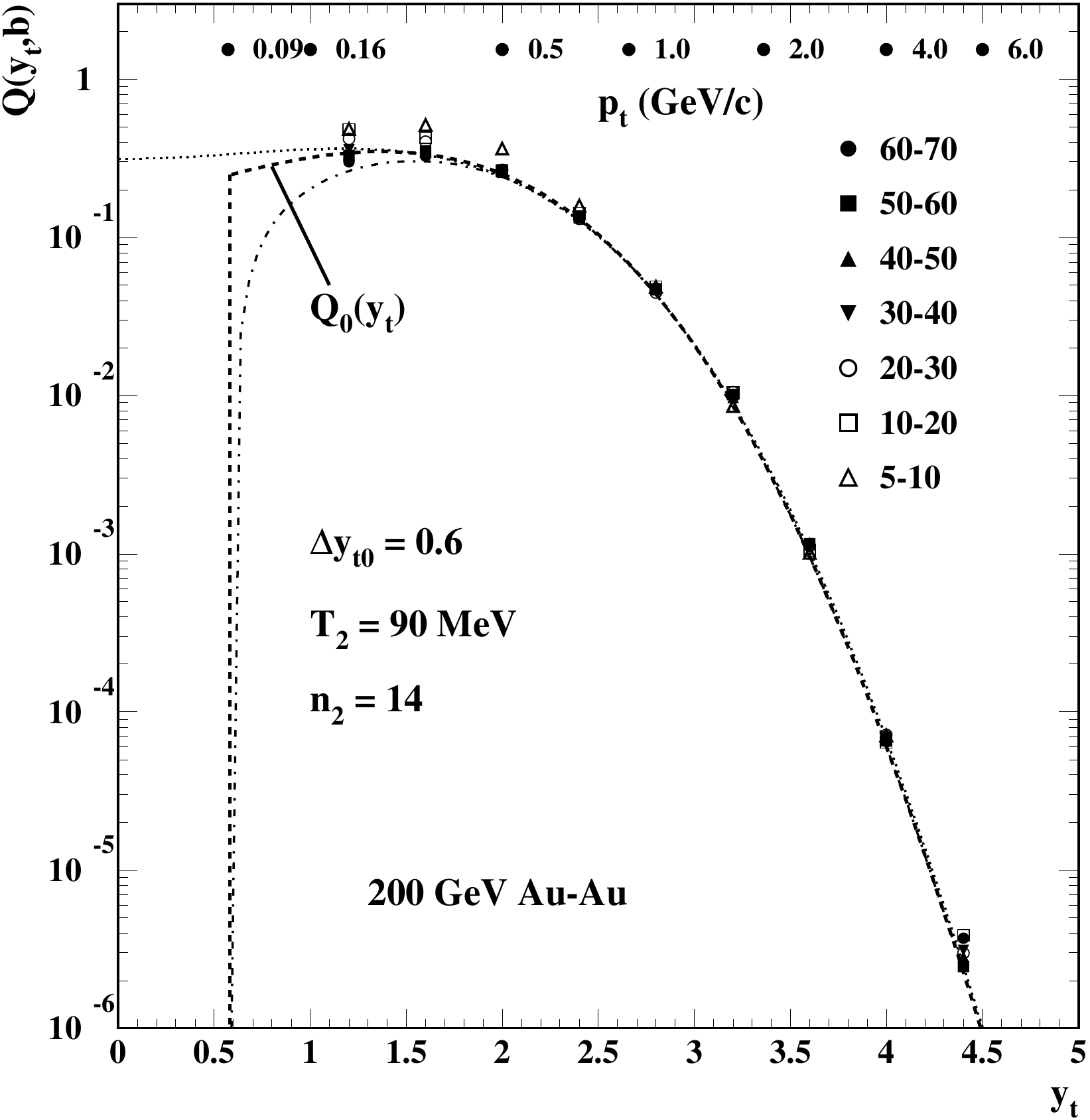}
\caption{\label{quadspec}
Unit-normal quadrupole spectra for seven centralities of 200 GeV \auau\ collisions derived from Eq.~(\ref{qb}). The spectrum shapes are independent of \auau\ centrality: all $Q(y_t,b)$ coincide within systematic uncertainties.  The common shape denoted by $Q_0(y_t)$ is well described by a boosted L\'evy distribution with parameters $T_2$ and $n_2$.~\cite{quadspec}.
} 
 \end{figure}

Figure~\ref{quadspec} shows $Q(y_t,b)$ data for seven centrality bins of 200 GeV \auau\ collisions derived from pair ratios $B_Q(y_t,b)$ obtained in the present analysis (Fig.~\ref{data1}). 2D model fits with the AS dipole peak model are preferred because those fits are more stable. The $B_Q$ data are combined with SP spectra $\rho_0(y_t,b)$ and yields $\rho_0(b)$ from Ref.~\cite{hardspec} to compute $Q(y_t,b)$. 
The ratio is undefined for the 0-5\% bin since $v_2\{\text{2D}\}(y_t,b)$ for that centrality is consistent with zero for $y_t > 2$.  We observe that within the data uncertainties $y_t$-differential quadrupole data follow a universal spectrum shape above $y_t = 2$ ($p_t = 0.5$ GeV/c) represented by unit-normal quadrupole spectrum $Q_0(y_t)$ (dashed curve) previously derived from MB PID $v_2$ data in the form of a {\em boosted L\'evy distribution}~\cite{quadspec}. $Q_0$ is {\em not a fit} to data from the present analysis. We conclude that quadrupole source boost $\Delta y_{t0} \approx 0.6$ for unidentified hadrons (mainly pions) is approximately {\em independent of \auau\ centrality}. That result is consistent with 0-10\% Lambda $v_2(p_t)$ data from Ref.~\cite{newlam}.

There are currently no accurate parametrizations available for 62 GeV SP spectra $\rho_0(y_t,b)$. Thus, complete reconstruction of quadrupole spectra for the lower energy is not possible. Based on the results in Fig.~\ref{boost2} the same quadrupole spectrum universality 
may persist there. 

In Fig.~\ref{boost2} we observed that the plotted data for the three most-central bins in the interval $y_t > 2$ ($p_t > 0.5$ GeV/c) are suppressed relative to the trend for other centralities. 
In Fig.~\ref{quadspec} those data have been rescaled by factors $g(b)$ in Eq.~(\ref{qb}). The resulting vertical shifts bring the spectra into coincidence above $y_t = 2$. The agreement of all quadrupole spectrum shapes with common form $Q_0(y_t)$ over that interval is within data uncertainties. The $y_t$-integral quantity $v_2\{\text{2D}\}(b)$ from Ref.~\cite{davidhq} is consistent with the $v_2\{\text{2D}\}(y_t,b)$ values  below $y_t = 2$ for all centralities (because of the SP spectrum shape), but a substantially different $v_2'(b)$ describes the $v_2(y_t,b)$ centrality trend above $y_t = 2$ for more-central collisions.


Results from Ref.~\cite{quadspec} and Fig.~\ref{quadspec} taken together suggest that all quadrupole spectra for any \aa\ centrality and for any hadron species follow universal $Q_0(m_t')$ in the boost frame except for most-central \auau\ collisions where an additional reduction factor $g(b) < 1$ is required. That conclusion may be contrasted with arguments for quark-number scaling of $v_2(p_t)$ data summarized in Ref.~\cite{quadspec}.  


\section{ $\bf v_2(y_t,b,\sqrt{s_{NN}})$ Parametrization} \label{v2ytparam}

Model-parameter trends derived from 2D model fits to $y_t$-differential histograms  reveal factorization of quadrupole systematics on centrality and hadron $y_t$, the factors described by simple functions.

Combining above results the $y_t$-differential quadrupole amplitude can be described in factorized form as
\bea \label{v2ytfac}
V_2^2\{\text{2D}\}(y_t,b) &\approx& \langle 1/p_t \rangle(b) V_2^2\{\text{2D}\}(b)\, p_t Q_0(y_t).
\eea
That simplicity becomes apparent only in  terms of extensive measure $V_2^2(y_t,b)$ obtained by eliminating the SP spectrum in the denominator of intensive measure $v_2^2(y_t,b)$ with its strong jet contribution. 
The simplicity of Eq.~(\ref{v2ytfac}) is unique to the quadrupole spectrum.


The quadrupole amplitude determined in this study is
\bea \label{bq2d}
B_Q\{\text{2D}\}(y_t,b) &=& v_2^2\{\text{2D}\}(y_t,b)  \\ \nonumber
&=& v_2\{\text{2D}\}(b) \,v_2\{\text{2D}\}(y_t,b),
\eea
and we use the $v_2\{\text{2D}\}(b)$ parametrization from Refs.~\cite{davidhq,noelliptic} (Sec.~\ref{ytintsyst}) to infer $v_2\{\text{2D}\}(y_t,b)$. 
Based on Sec.~\ref{quadspecc} the $v_2\{\text{2D}\}(y_t,b)$ data can be represented accurately by a simple parametrization, just as for the $y_t$-integral case. From above we have the relation $V_2^2(y_t,b) = \rho_0(b) \rho_0(y_t,b) v_2^2(y_t,b)$. 
Rearranging Eq.~(\ref{v2ytfac}) accordingly we obtain the parametrization
\bea \label{universalv2}
v_2\{\text{2D}\}(p_t,b) &=& p_t \langle 1/p_t \rangle v_2\{\text{2D}\}(b)  \left\{\frac{Q_0(p_t)}{\rho_0(p_t,b)/\rho_0(b)}\right\}  \nonumber \\
\frac{Q_0(p_t)}{\rho_0(p_t,b)/\rho_0(b)} &\approx&\exp(-p_t / P) \times f(y_t,b),
\eea
which can be compared with $v_2(p_t,b)$ data. 
The ratio of unit-normal spectra in curly brackets is approximated with reasonable accuracy by $\exp(-p_t/P)$. An $O(1)$ empirical factor $f(y_t,b)$ representing deviations from that function for more-central collisions is
\bea
f(y_t,b) \hspace{-.03in} &=& \hspace{-.03in} 1 + C(b)\left[  \text{erf}(y_t - 1.2) - \text{erf}(1.8 - 1.2) \right] \\ \nonumber
C(b) &=& 0.12 - (\nu - 3.4)/5 - [(\nu - 3.4)/2]^5.
\eea
Function $f(y_t,b)$ decreases with $y_t$ from values exceeding 1 for $y_t < 2$ to a constant value $g(b) \leq 1$ for $y_t > 2$. The curve crosses through unity near $y_t = 1.8$ ($p_t \approx 0.4$ GeV/c). Values $g(b) =$ 0.9, 0.75, 0.55 are inferred from the data for centrality bins 20-30\% through 5-10\% (the 0-5\% $B_Q$ data for $y_t > 2$ are consistent with zero).

 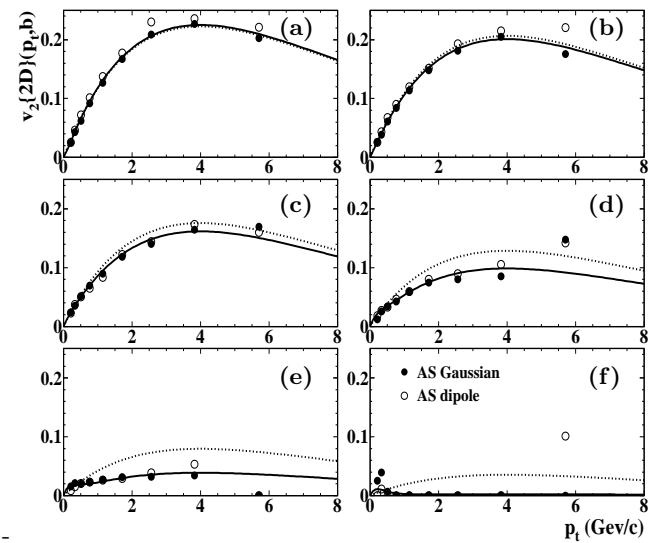
\begin{figure}[h]
-  \includegraphics[width=3.3in,height=2.8in]{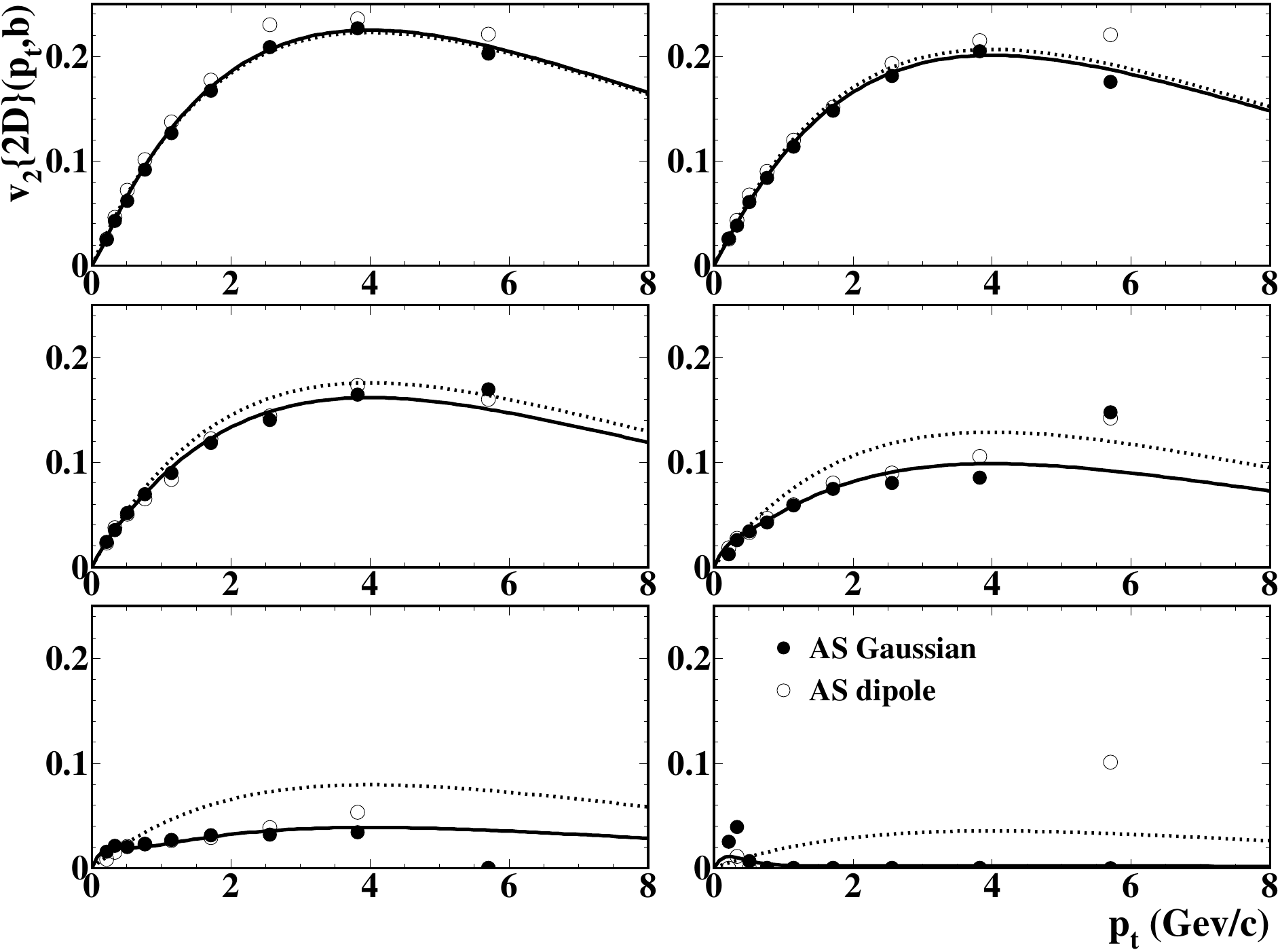}
\put(-140,192) {\bf (a)}
 \put(-140,125) {\bf (c)}
 \put(-140,60) {\bf (e)}
\ \put(-23,192) {\bf (b)}
 \put(-23,125) {\bf (d)}
 \put(-23,60) {\bf (f)}
\caption{\label{param}
$p_t$-differential $v_2\{\text{2D}\}(p_t,b)$ trends from 200 GeV \auau\ collisions (points) compared to a parametrization of $v_2\{\text{2D}\}(p_t,b)$ data given by Eq.~(\ref{universalv2}) (solid curves)
for (a) 40-50\%, (b) 30-40\%, (c) 20-30\%, (d) 10-20\%, (e) 5-10\%, (f) 0-5\% centralities. Solid points are for the AS 1D Gaussian peak model and open circles are for the AS dipole peak model in Eq.~(\ref{estructfit}). Dotted curves are Eq.~(\ref{universalv2}) without factor $f(y_t,b)$.
} 
 \end{figure}


Figure~\ref{param} shows comparisons between Eq.~(\ref{universalv2}) and 200 GeV data from the present $y_t$-differential analysis (points).
The solid dots and open circles represent $v_2\{\text{2D}\}\{p_t,b\}$ data from 2D fits with AS 1D Gaussian and AS dipole peak models respectively. The parametrization of Eq.~(\ref{universalv2}) (solid curves) describes the $\{\text{2D}\}$ data accurately over a large kinematic range. Significant differences arising from the choice of AS 1D peak model appear only for large $y_t$ values as expected.
Thus, the algebraic relations in Refs.~\cite{davidhq,noelliptic} and Eq.~(\ref{universalv2}) are confirmed by comparisons with these $v_2\{\text{2D}\}(y_t,b)$ data.

The dotted curves exclude the factor $f(y_t,b)$ and therefore have the simple form $\propto p_t \exp(-p_t/P)$ with $P = 4$ GeV/c for 200 GeV. The more-peripheral data (possibly down to \nn\ collisions) are consistent with that parametrization (follow the dotted curves). NJ quadrupole data for three more-central bins, and for $y_t > 2$ ($p_t > 0.5$ GeV/c), fall increasingly below the trend predicted by the parametrization of $v_2\{\text{2D}\}(b)$ from Ref.~\cite{davidhq}, and quadrupole data for 0-5\% central collisions are consistent with zero throughout that interval.



\section{Systematic Uncertainties} \label{uncert}



Statistical and systematic uncertainties are discussed for $y_t$-differential $B_Q$ data and for inferred quadrupole spectrum trends.
For this differential study the choice of fit model is a compromise between minimizing systematic errors and employing the same model to cover large kinematic intervals on $y_t$ and centrality $\nu$. For this discussion we refer to the model elements in Eq.~(\ref{estructfit}). 

\subsection{2D fit-model elements}

At lower $y_t$ the 2D exponential (BEC + electrons) peak narrows with increasing centrality while the SS 2D peak broadens on $\eta$, insuring accurate distinctions. At higher $y_t$ the exponential peak amplitude drops rapidly to zero. Thus, except for peripheral collisions that element can be dropped from the 2D fit model. To minimize systematic fit errors from that source the central three bins are omitted from all 2D fits (bin errors greatly increased).

At small $\nu$ (peripheral collisions, below the sharp transition) the SS 2D peak is narrow on $\eta$ and $\phi$ and accurately described by a 2D Gaussian for all $y_t$ bins. In more-central \auau\ collisions the SS 2D peak broadens on $\eta$ but remains narrow on $\phi$. For most $y_t$ bins the 2D peak is still accurately described by a 2D Gaussian. At larger $y_t$ the peak is distorted on $\eta_\Delta$, developing non-Gaussian tails as observed in trigger-associated analysis~\cite{trigger,starridge}. The 2D Gaussian model, while no longer bin-wise accurate, does estimate the peak amplitude and r.m.s.\ widths satisfactorily for the present study focusing on the azimuth quadrupole component.


In more-peripheral collisions (and in the $y_t$-integral analysis) the AS 1D peak is broad enough to be modeled accurately by a single dipole term. In more-peripheral collisions and at larger $y_t$ the AS 1D peak appears to narrow. 
If the width of the AS peak becomes substantially smaller than $\pi / 2$ and the AS dipole model is employed the quadrupole component of the AS 1D peak might appear as a bias in the inferred quadrupole amplitude. That bias source can be investigated by replacing the AS dipole by an AS 1D Gaussian model and refitting the 2D data. Any differences in inferred $B_Q\{\text{2D}\}(y_t,b)$ establish the systematic uncertainty from that source in the inferred NJ azimuth quadrupole amplitude, typically at the few-percent level as with other fit uncertainties.


\subsection{2D fit-quality systematics}

\begin{figure}[h]
  \includegraphics[width=1.65in,height=1.5in]{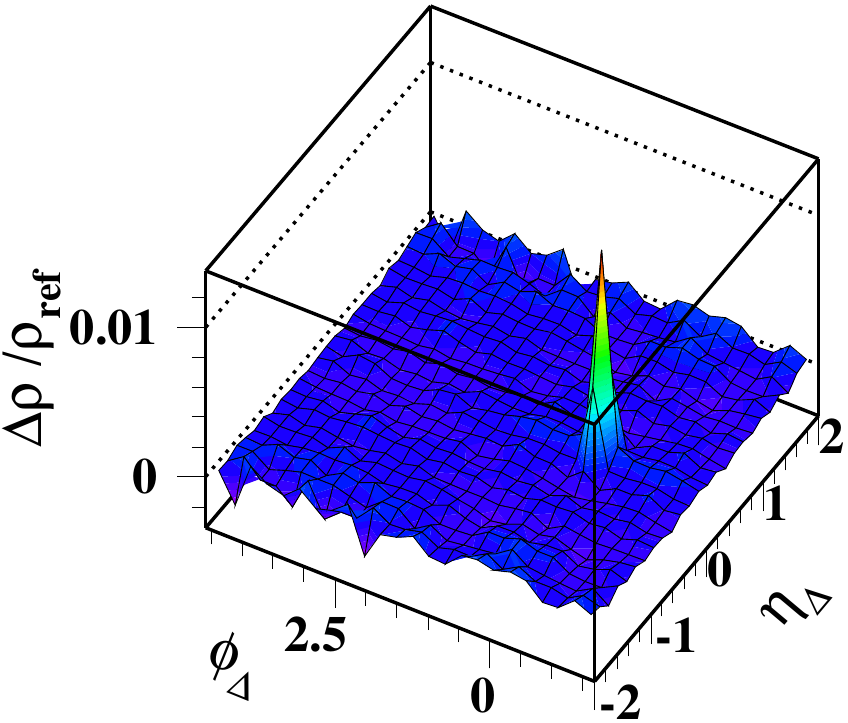}
\includegraphics[width=1.65in,height=1.5in]{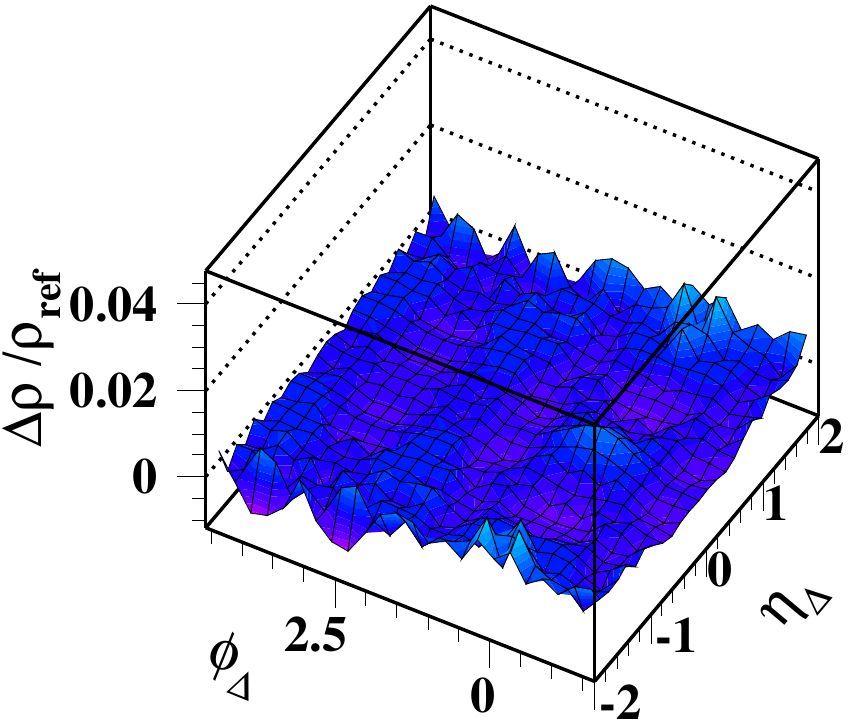} 
 \caption{\label{resids2} (Color online)
 Residuals from 2D fits to $y_t$-differential histograms from 40-50\%-central 200 GeV \auau\ collisions for
Left:  $y_t \in [1.8,2.2]$ ($p_t \in$ [0.4,0.65] GeV/c) and
Right:   $y_t \in [3.4,3.8]$ ($p_t \in$ [2,3.3] GeV/c). The vertical scales are half the range of the corresponding panels in Fig.~\ref{ptdiff1}.
}  
 \end{figure}

Figure~\ref{resids2} shows typical fit residuals for lower- and higher-$y_t$ bins using the eight-parameter fit model. For lower-$y_t$ bins (left panel) the residuals  do not contain significant large-scale structure and are generally consistent with statistical uncertainties. 
The  $\chi^2/ndf \approx 0.8 - 2.5$. The BEC + electron peak at the origin appearing in the residuals (not described by the fit model) does not contribute to the $\chi^2$ because those three bins are excluded from the fit (assigned artificially large errors). In the higher-$y_t$ bin the BEC + electron peak is negligible, consistent with the $p_t$ dependence of both mechanisms.


To summarize systematic uncertainty trends  for the $y_t$-differential analysis we identify three zones on kinematic space $(y_t,\nu)$. Zone A is more-peripheral collisions ($\nu < 2$, $1-\sigma / \sigma_0 \leq 0.3$) for all $y_t$, zone B is more-central collisions ($\nu > 2$) for lower $y_t$, and zone C is more-central collisions for higher $y_t$. The $y_t$ boundary between zones B and C is approximately $y_t = 3.8$ ($p_t \approx 3$ GeV/c). In zone A the eleven-parameter model function for $y_t$-integral analysis from Refs.~\cite{anomalous,davidhq} is required for satisfactory fits. In zone B either the eleven- or eight-parameter model function provides similar fit quality. 

In zone C the eleven-parameter model is excluded because of fitting ambiguities between part of the SS 2D peak structure and the 2D exponential model element, but the simpler eight-parameter model function provides an adequate description when three bins near the origin are excluded from the fit. As noted, the simpler model function is therefore utilized to cover zones B and C reported in this $y_t$-differential analysis, and zone A is not reported. Systematic uncertainties in zone B are small and consistent with statistical and fit errors. Uncertainties in C may be significant and are explicitly estimated. 

Figure~\ref{resids2}  (right panel) shows small but significant residuals structure resulting from the non-Gaussian shape of the SS 2D peak appearing in larger-$y_t$ bins (zone C): excesses at the origin and near the acceptance boundaries on $\eta_\Delta$ for $\phi_\Delta \approx 0$. A small excess in the {\em inferred} quadrupole $B_Q$ is also observed (e.g., depression near $\phi_\Delta = \pi$) due to the non-Gaussian SS peak shape. The peak-peak residual quadrupole amplitude is about 0.001 for $B_Q \approx 0.014$ [Fig.~\ref{data1} (d)]. Thus, the relative uncertainty is $\approx 0.0005 / 0.014 \leq $ 5\%, comparable to the typical fit uncertainties there. The inferred quadrupole data are generally stable against minor changes in jet-related fit model elements or the SS 2D peak shape, except for the largest $y_t$ values where a substantial systematic uncertainty (20\%) must be assigned to the $B_Q$ data. What matters most for the NJ quadrupole is simply the presence in the fit model of a SS 2D peak narrow on azimuth.



\subsection{Quadrupole spectrum uncertainties}



The SP spectrum parametrization from Ref.~\cite{hardspec} is not constrained by data below $y_t = 2$. Thus, some systematic deviations from the $Q_0(y_t)$ reference may  be due to inaccuracies in the modeled SP spectrum structure. Also, we have approximated the hadron spectrum for this study by the pion spectrum alone. Protons and kaons do play a significant role in the hadron spectrum shape, and those spectra are substantially different from the pion spectrum, introducing a further source of systematic bias.

 \begin{figure}[h]
  \includegraphics[width=3.in,height=1.5in]{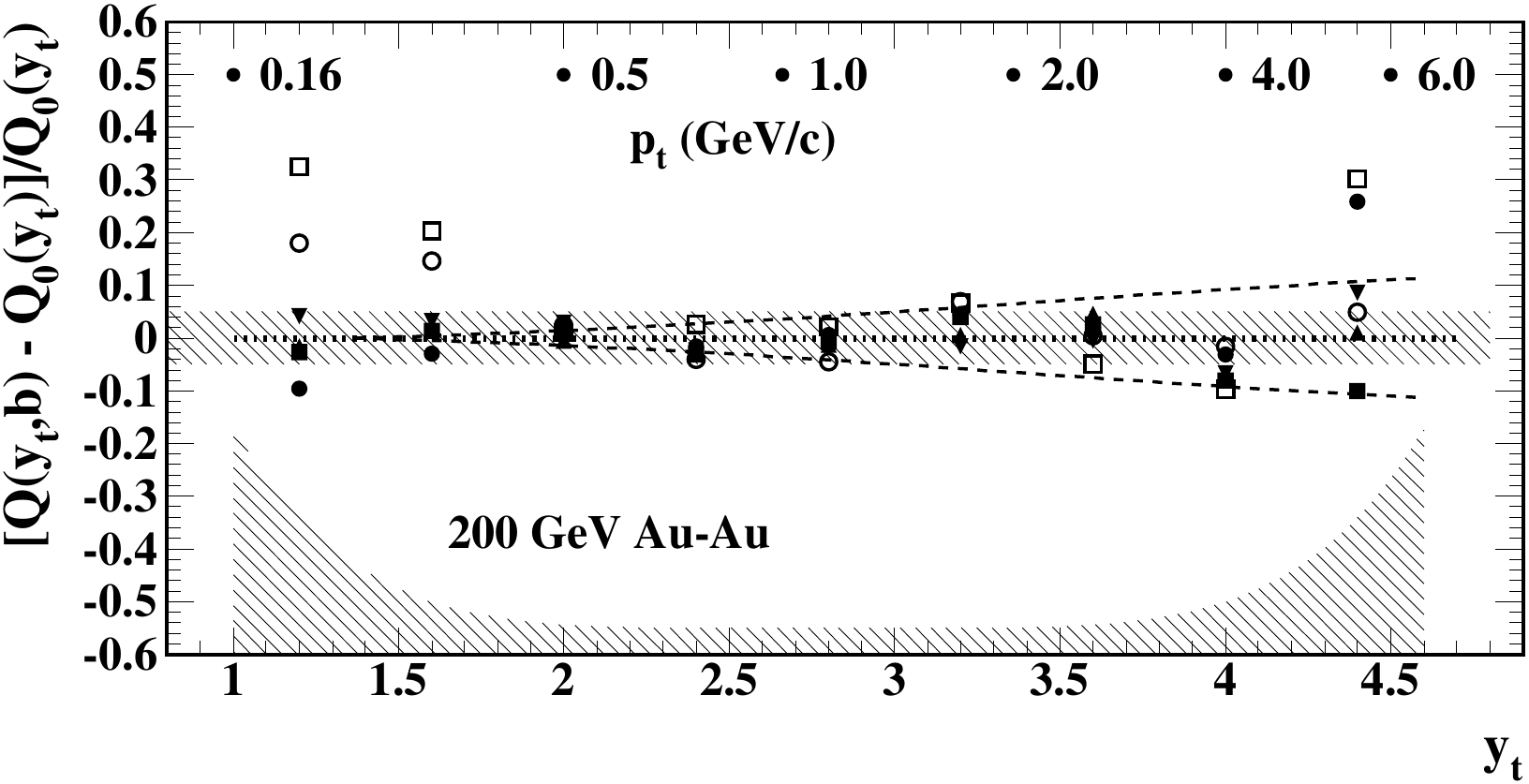}
 \caption{\label{resids3} 
Data-model deviations $Q(y_t,b) - Q_0(y_t)$ are shown relative to fixed spectrum model $Q_0(y_t)$. The upper hatched band indicates that the r.m.s. data deviation is less than 5\% except near the acceptance endpoints. There are systematic deviations below 0.5 GeV/c and larger fitting uncertainties for the largest $y_t$ bin. The lower hatched band provides an estimate of systematic uncertainties above $y_t = 1.8$ and the range of variations corresponding to $v_2(b)$ vs $v_2'(b)$ below $y_t = 1.8$. The dashed curves are explained in the text.
} 
\end{figure}

Figure~\ref{resids3} shows relative deviations of unit-normal quadrupole spectrum data $Q(y_t,b)$ from universal spectrum reference $Q_0(y_t)$. The symbols are defined as in Fig.~\ref{quadspec}. The more-central data are scaled by factor $g(b) < 1$ as in Fig.~\ref{quadspec} so that spectrum shapes for $y_t > 2$ ($p_t > 0.5$ GeV/c) can be compared. The more-central data then exhibit large deviations below $y_t = 2$ due to the rescaling. Such deviations could represent variations in SP spectra below $y_t = 2$, but they may also reflect significant changes in the actual quadrupole spectra with physical implications. The deviations above $y_t = 4$ are consistent with fit instabilities and sparse statistics. However, we also expect an excess at larger $y_t$ because those data are derived using the AS dipole model. If the AS 1D peak narrows at larger $y_t$ (expected) the quadrupole component of the AS 1D peak may then contribute a positive bias to the inferred $v_2\{\text{2D}\}$ data.

We can establish an upper limit on possible source boost variations with a Taylor expansion of $Q(y_t,b)$ about $Q_0(y_t)$ relative to variations in boost $\Delta y_{t0}$. The two dashed curves represent $\pm \delta y_{t0}\, d\log[Q_0(y_t)]/dy_t$, with $\delta y_{t0} = 0.02 \Delta y_{t_0} = 0.012$ or 2\% of the mean source boost. The data lie well within those limits for $y_t > 1.8$ ($p_t > 0.4$ GeV/c). 
Boost variations with centrality are comparable to or smaller than those observed small deviations on $y_t$.
The comparison suggests that the mean quadrupole source boost does not change by more than a few percent over a broad centrality interval.
In general, the $Q_0(y_t)$ universal quadrupole spectrum represents the $Q(y_t,b)$ data well over $p_t$ in 0.35-4 GeV/c ($y_t $ in 1.8-4).

 \section{Comparisons among $\bf v_2$ methods} \label{methcomp}

Substantial differences appear between the NJ azimuth quadrupole $v_2\{\text{2D}\}$ derived from model fits to 2D angular correlations and $v_2\{\text{method}\}$ data derived from conventional NGNM  (equivalent to 1D model fits to azimuth correlations)~\cite{davidhq,davidhq2}. In this section we provide detailed comparisons among several methods and consider possible sources of observed differences.


\subsection{Algebraic relation between $\bf v_2\{2\}$ and $\bf v_2\{\text{EP}\}$}



To establish the relation between conventional event-plane $\{\text{EP}\}$ and two-particle correlation $\{2\}$ methods we first note that $v_2\{\text{EP}\}$ is defined by~\cite{poskvol,flowmeth}
\bea
v_2\{\text{EP}\} = \frac{v_{2,obs}}{\langle \cos[2(\Psi_2 - \Psi_r)] \rangle},
\eea
where the numerator is ``observed'' $v_{2,obs}$ (defined below), and the denominator is described as the {\em event-plane resolution}.
To relate $v_{2,obs}$ to $v_2\{2\}$ we define the $m=2$ (azimuth quadrupole) $Q$ vector by~\cite{complex,2002,flowmeth}
\bea \label{qvec}
\vec Q_2 &=& \frac{1}{2\pi \Delta \eta} \sum_{i=1, \in \Delta \eta}^n \vec u(2\phi_i) \equiv Q_2 \vec u(2 \Psi_2),
\eea
with unit vectors $\vec u(\phi)$ and {\em event-plane} angle $\Psi_2$. The vector notation is an alternative to that in Ref.~\cite{complex} based on complex quantities. $\vec Q_2$ as defined in Eq.~(\ref{qvec}) is a 2D angular density. We then have
\bea \label{q22}
Q_2^2 &=& \vec Q_2 \cdot \vec Q_2 = \frac{\rho_0}{2\pi \Delta \eta} + V_2^2\{2\}
\eea
as in Eq.~(\ref{bigv2}) but {\em with self pairs included} in the first term on the RHS. It is notable that $V_2^2\{2\}$ and $Q_2^2$ differ only by the self-pair term.
We then have
\bea \label{v2obs}
v_{2,obs} &\equiv& \langle \cos[2(\phi - \Psi_2)] \rangle \\ \nonumber
&=& \frac{1}{n} \sum_{i=1}^{n} \vec u(2\phi_i) \cdot \vec u(2\Psi_{2,i}), ~~~\text{or} \\ \nonumber
Q'_2 v_{2,obs} &=& \frac{1}{2\pi \Delta \eta} \frac{1}{n} \sum_{j \neq i = 1}^{n, n-1} \vec u(2\phi_i) \cdot \vec u(2\phi_j) \\ \nonumber
&=&  v_2\{2\} V_2\{2\},
\eea
where $Q'_2 = Q_2\sqrt{(n-1)/n}$, and $X_{2,i}$ indicates that the $i^{th}$ term is excluded from a sum over $j$. The summation condition $j \neq i$ in Eq.~(\ref{v2obs}) (third line) excludes self-pairs from that pair sum but {\em not} from $Q_2$ in Eq.~(\ref{q22}) (or $Q_2'$). From Ref.~\cite{flowmeth} we obtain the ``resolution'' measure
\bea \label{epres}
\overline{ \cos^2[2(\Psi_2 - \Psi_r)]} &=&   \frac{ \overline{n\, V_2^2\{2\}}}{\overline{ (n-1)\,  Q_2^2}} \\ \nonumber
&\approx& {\frac{n  v_2^2\{2\}}{1 + n v_2^2\{2\}}},
\eea
where $n  v_2^2\{2\}$ serves as a statistical figure of merit for ratio $v_2$ analogous to $\sigma p^2$ ($\sigma$ is a nuclear cross section) or $Lp^2$ ($L$ is a beam luminosity) for measurements of polarization ratio $p$. 
The $\{\text{EP}\} \leftrightarrow \{2\}$ relation is then
\bea
v_2\{2\} \hspace{-.005in} &=& \hspace{-.005in} v_{2,obs}\frac{Q_2}{V_2\{2\}}\hspace{-.002in} \sqrt{\frac{n-1}{n}}  \\ \nonumber
&=& \hspace{-.002in} \frac{v_{2,obs}}{\langle \cos[2(\Psi_2 - \Psi_r)] \rangle_\text{r.m.s.}}\hspace{-.002in} \approx\hspace{-.02in} v_2\{\text{EP}\}.
\eea
Small $\{\text{EP}\} \leftrightarrow \{2\}$ differences may arise from covariances corresponding to non-Poisson multiplicity fluctuations. The ``event-plane resolution'' correction is required because invocation of an event-plane estimate via $\vec Q_2$ implicitly includes a self-pair contribution. Excluding self pairs in Eq.~(\ref{v2obs}) {does not} remove the $Q_2$ bias. The $v_2\{\text{EP}\}$ estimate does not necessarily relate to an \aa\ reaction plane---it represents all 2D correlation structure {\em including jets}. We hereafter refer exclusively to $v_2^2\{2\}$ or $V_2^2\{2\}$ except when introducing published $v_2\{\text{EP}\}$ data.


\subsection{Algebraic relation between $\bf v_2\{2\}$ and $\bf v_2\{\text{2D}\}$}

The quadrupole power spectrum element $V_2^2\{2\}$ (equivalent to $ \rho_0 A_Q\{2\},~ \rho_0^2 B_Q\{2\}$) represents the total azimuth quadrupole component for all angular correlations, including both jet-related structures and any nonjet structure that might be identified with flows. As noted in Sec.~\ref{modelmotives} the $\eta_\Delta$ dependence of 2D angular correlations can be employed to separate unique correlation components via 2D model fits, as in Refs.~\cite{axialci,anomalous} and the present analysis. For almost all collision conditions we observe that the AS structure of 2D angular correlations is uniform on $\eta_\Delta$ within $|\eta| < 1$ and completely described by a NJ azimuth quadrupole represented by $A_Q\{\text{2D}\}$ and  an AS dipole component.

The only remaining nontrivial structure observed in more-central \auau\ collisions is a SS  2D peak (consistent with intrajet correlations). Because the AS dipole is orthogonal to all other multipoles the SS 2D peak is the only other significant contributor to total quadrupole $V_2^2\{2\} = \rho_0^2 v_2^2\{2\}$ {\em in more-central \aa\ collisions}.  The SS 2D peak per-particle quadrupole amplitude (Fourier coefficient) is given by~\cite{multipoles}
\bea \label{sseq}
2 A_Q\{\text{SS}\}(b) &=& F_2(\sigma_{\phi_\Delta}) G(\sigma_{\eta_\Delta};\Delta \eta) A_{2D},
\eea
where $A_{2D} = \rho_0 B_{2D}$ is the per-particle amplitude of the fitted SS 2D peak with r.m.s.\ widths $(\sigma_{\eta_\Delta},\sigma_{\phi_\Delta})$, $F_2$ is the $m=2$ Fourier component of a unit-amplitude 1D Gaussian on azimuth with width $\sigma_{\phi_\Delta}$
\bea \label{fm}
2F_m(\sigma_{\phi_\Delta}) &=& \sqrt{2/\pi}\, \sigma_{\phi_\Delta} \exp\left( - m^2 \sigma_{\phi_\Delta}^2 / 2\right),
\eea
and $G \leq 1$ is a calculated $2D \rightarrow 1D$ $\eta$ projection factor defined in Ref.~\cite{multipoles}. 
We thus obtain the relation
\bea \label{aqsum}
A_Q\{2\} &=& A_Q\{\text{2D}\} + A_Q\{\text{SS}\}
\eea
plus a small contribution from BEC + electron pairs in more-peripheral collisions.
Jet-related quadrupole $A_Q\{\text{SS}\}$ may be identified with ``nonflow''~\cite{gluequad,tzyam,multipoles}. Nonjet quadrupole $A_Q\{\text{2D}\}$ would correspond to elliptic flow if that phenomenon is relevant. We test that relation with results from the present and previous 2D correlation analysis and published $v_2\{\text{method}\}$ data in the next subsection.

Strategies have been adopted to reduce nonflow (mainly jet contributions) to $v_2$ by excluding some parts of the nominal $(\eta_1,\eta_2)$ acceptance from NGNM calculations~\cite{multipoles}. For instance, some $\eta_\Delta$ interval centered at zero may be excluded from projections onto $\phi_\Delta$ by ``estimating the reaction plane'' with large-$\eta$ detectors~\cite{2004,cpanp,alicev2}.  The motivation is exclusion of jet-related structure $A_Q\{\text{SS}\}$ from azimuth projections $A_Q\{2\}$ {\em based on assumptions} about the jet fragment distribution on $\eta$.

Such $\eta$ pair cuts may be less effective at distinguishing jet-related structure from a NJ quadrupole than 2D model fits applied within a more-limited $\eta$ acceptance. In more-central \auau\ collisions the SS 2D peak is strongly elongated and may develop non-Gaussian tails extending over a large $\eta_\Delta$ interval~\cite{trigger}. The effects of $\eta$-exclusion cuts are then quite uncertain and may have little impact on jet-related biases in $v_2\{\text{method}\}$ data~\cite{multipoles}.

\subsection{$\bf y_t$-differential data comparisons} \label{ytcomp} 

\begin{figure}[h]
  \includegraphics[width=1.65in,height=1.65in]{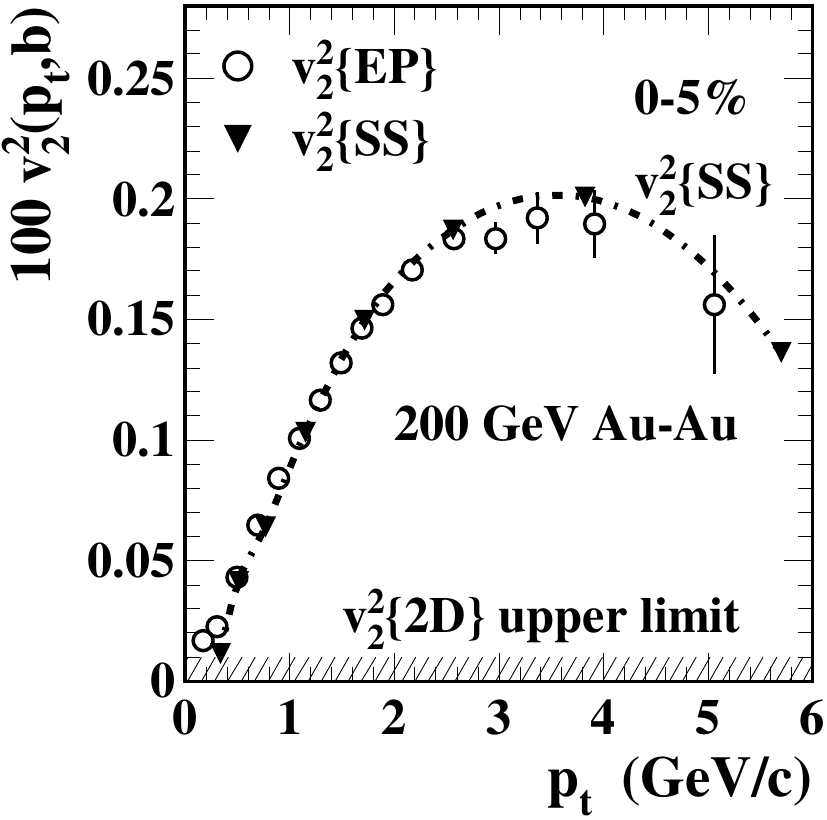}
  \includegraphics[width=1.65in,height=1.68in]{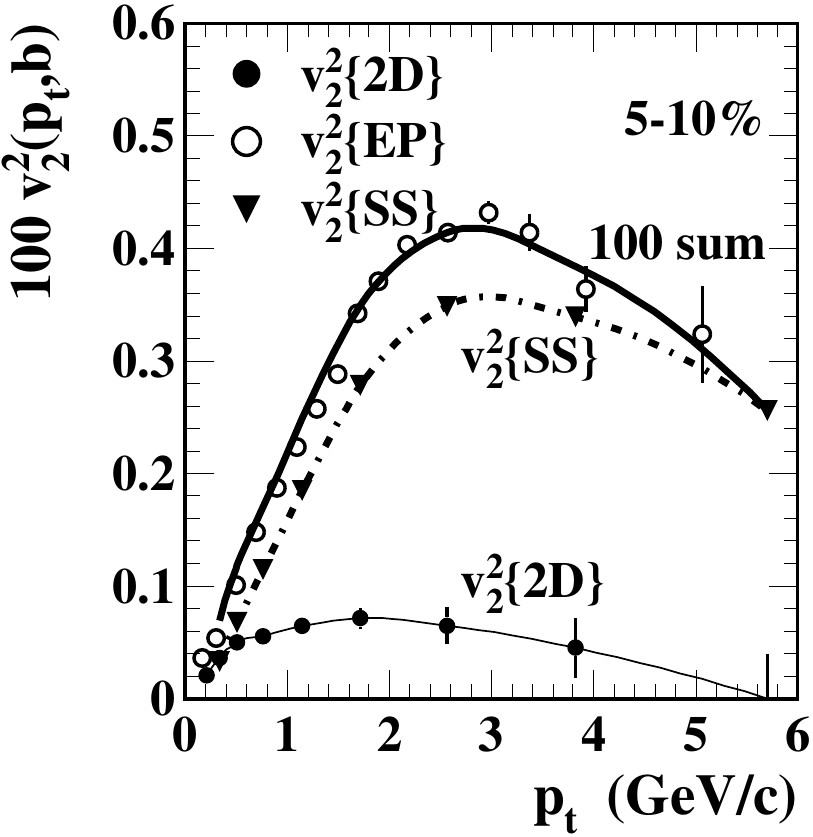}
\caption{\label{ytcomp1}
Left: Comparison of measured 2D angular correlations represented by $v_2^2\{\text{SS}\}(p_t,b)$ (dash-dotted curve) and $v_2^2\{\text{2D}\}(p_t,b)$ (hatched region, upper limit) with published $v_2^2\{\text{EP}\}(p_t,b)$ data (open circles) for 0-5\% central 200 GeV \auau\ collisions.
Right: Similar comparison for 5-10\% central \auau\ collisions showing  the close correspondence between the sum $v_2^2\{\text{2D}\}(p_t,b) + v_2^2\{\text{SS}\}(p_t,b)$ (bold solid curve) and published $v_2^2\{\text{EP}\}(p_t,b)$ data (open circles) as in Eq.~(\ref{aqsum}). The light solid curve through the $v_2^2\{\text{2D}\}$ data (solid dots) represents the parametrization in  Eq.~(\ref{universalv2}).
} 
 \end{figure}

Figure~\ref{ytcomp1} shows published $v_2^2\{\text{EP}\}(p_t,b)$ data (open circles) compared to $v_2^2\{\text{2D}\}(p_t,b)$ data from the present analysis (solid points or hatched upper limit) and ``nonflow'' prediction $v_2^2\{\text{SS}\}$ derived from characteristics of the SS 2D peak measured in this analysis. The $v_2^2\{\text{EP}\}(p_t,b)$ points are obtained by combining $v_2(b)$ and $v_2(p_t,b)$ measurements from Ref.~\cite{2004}. The hatched region in the left panel denotes an upper limit on $v_2^2\{\text{2D}\}$ [compare with Fig.~\ref{param} (f)]. The bold solid curve in the right panel is defined (without the factor 100) by $v_2^2\{2\} = v_2^2\{\text{2D}\} + v_2^2\{\text{\text{SS}}\}$ per Eq.~(\ref{aqsum}).

We find that the measured $v_2^2\{\text{EP}\} \approx v_2^2\{2\}$ trend is predicted by a combination of $v_2^2\{\text{2D}\}$ data and $v_2^2\{\text{SS}\}$ representing the $m = 2$ Fourier component of the SS 2D jet peak projected onto 1D azimuth. The dash-dotted curves $v_2^2\{\text{SS}\}(p_t,b)$ derived from SS 2D peak properties inferred from this analysis can be interpreted as the jet contribution to $v_2^2\{2\}$. We confirm the trend $v_2^2\{2\} = v_2^2\{\text{SS}\} + v_2^2\{\text{2D}\}$ for $y_t$-differential data based on the detailed $\eta$ dependence of 2D angular correlations. There is no adjustment to accommodate the $v_2^2\{\text{EP}\}$ data.

\begin{figure}[h]
  \includegraphics[width=1.65in,height=1.68in]{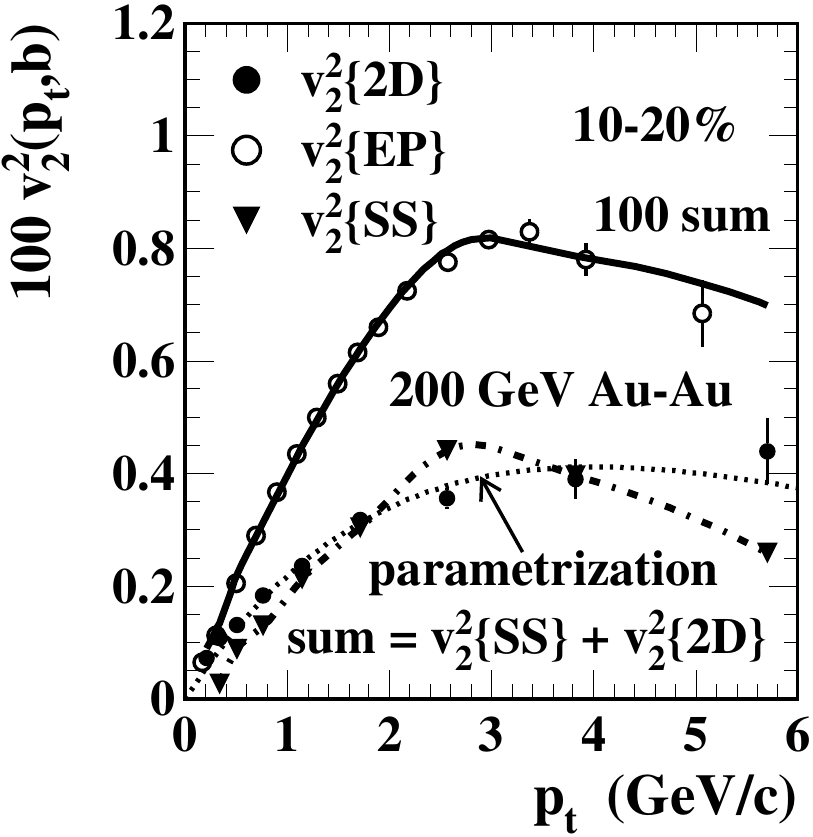}
  \includegraphics[width=1.65in,height=1.65in]{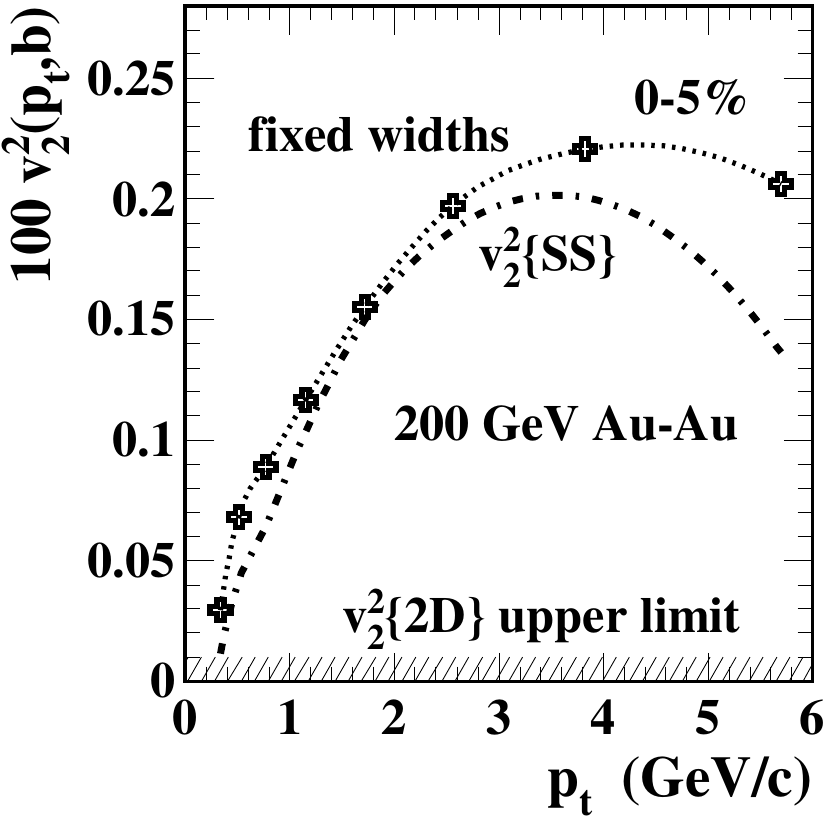}
\caption{\label{ytcomp2}
Left: Comparison similar to Fig.~\ref{ytcomp1} for 10-20\% central \auau\ collisions showing near-equal contributions from NJ quadrupole $v_2^2\{\text{2D}\}(p_t,b)$ and jet-related quadrupole $v_2^2\{\text{SS}\}(p_t,b)$. The bold solid curve is the sum of \{\text{2D}\} and \{\text{SS}\} data. The bold dotted curve is the parametrization of $v_2\{\text{2D}\}(p_t,b)$ from Eq.~(\ref{universalv2}) combined with a $v_2\{\text{2D}\}(b)$ value derived from Refs.~\cite{davidhq,noelliptic} (Sec.~\ref{ytintsyst}).
Right: The quadrupole amplitude of the SS 2D peak (dash-dotted curve) derived from Eq.~(\ref{sseq}) with data from Fig.~\ref{data1} [same as Fig.~\ref{ytcomp1} (left panel)] compared to a similar calculation using fixed SS 2D peak widths (dotted curve and points). The substantial differences illustrate the importance of accurately-measured SS 2D peak properties for understanding jet biases in $v_2\{2\} \approx v_2\{\text{EP}\}$ data .
} 
 \end{figure}

Figure~\ref{ytcomp2} (left panel) shows data for 10-20\% central \auau\ collisions including similar contributions from NJ quadrupole $v_2^2\{\text{2D}\}(p_t,b)$ and jet-related quadrupole $v_2^2\{\text{SS}\}(p_t,b)$. The sum (bold solid curve) accurately describes the published $v_2\{\text{EP}\}$ data. The parametrization of Eq.~(\ref{universalv2}) (dotted curve) describes the  $v_2^2\{\text{2D}\}(p_t,b)$ data over the entire $p_t$ acceptance.

Figure~\ref{ytcomp2} (right panel) illustrates the importance of accurate {jet-related} 2D correlation measurements. The dash-dotted curve is $v_2^2\{\text{SS}\}$ from Fig.~\ref{ytcomp1} (left panel) derived from Eqs.~(\ref{sseq}) and (\ref{fm}) based on measured SS 2D peak characteristics as in Fig.~\ref{data1}. The dotted curve and points represent the same computation with the SS 2D peak widths {\em held fixed} at $\sigma_{\eta_\Delta} = 2.5$ and $\sigma_{\phi_\Delta} = 0.65$ ($y_t$-integral values for that centrality). This exercise illustrates that accurate description of the $v_2^2\{\text{EP}\}$ data in Fig.~\ref{ytcomp1} (left panel) by the dash-dotted curve relies on full employment of measured SS 2D peak properties. 

We learn for instance that relative to the correct dash-dotted curve the dotted curve assuming fixed SS peak widths is too large at lower $p_t$ because the SS peak azimuth width is substantially larger there, leading to an overestimate of Fourier coefficient $F_2(\sigma_{\phi_\Delta})$ in Eq.~(\ref{fm}) by the fixed-width assumption. The dotted curve is too large at higher $p_t$ because the SS peak $\eta$-width reduction is not taken into account.  The substantial downturn in $v_2^2\{\text{EP}\}(p_t,b)$ for 0-5\% centrality at larger $p_t$ is solely due to strong narrowing of the SS 2D peak on $\eta_\Delta$ above 4 GeV/c {\em toward the \pp\ value}, as in Fig.~\ref{data1}  (b).
The combination of measured amplitude and widths of the SS peak from Fig.~\ref{data1} accurately describes the nonflow (jet) contribution $v_2^2\{\text{SS}\}$ to $v_2^2\{\text{EP}\} \approx v_2^2\{2\}$ (dash-dotted curve).

\section{Discussion} \label{discc}

We consider the implications  of differential $v_2\{\text{2D}\}(y_t,b)$ measurements from this analysis for conventional $v_2$ measurements and for interpretations of $v_2$ data in terms of hydrodynamic flows.  

\subsection{Fit models and interpretation of model elements}

We model \yt-differential 2D angular correlations with Eq.~(\ref{estructfit}) whose elements are motivated only by structures directly observed in the data, with no {\em a priori} physical assumptions. 
We then interpret the elements physically by comparison of data systematics with theoretical predictions. 
Based on measured trends for \pp\ and more-peripheral \aa\ collisions (95-50\% fractional cross section) we interpret the SS 2D peak and AS 1D peak elements as ``jet-related'' (see Sec.~\ref{nomen}). All other elements are  then referred to as ``nonjet,'' including the NJ azimuth quadrupole. 
The same terminology is retained in more-central \aa\ collisions although a jet interpretation may be questioned there. 
The ordered sequence---mathematical modeling {\em followed by} physical interpretation---is an essential feature.
The jet-related and non-jet terminology is complementary to flow-related and non-flow terms. However, jet phenomenology is well established from extensive HEP measurements and QCD theory whereas hydrodynamic flows in high-energy nuclear collisions (RHIC and LHC as opposed to the Bevalac/AGS energy regime) remain a matter of conjecture.

Jet-related and NJ quadrupole correlation components are separately identified. So-called nonflow bias is associated with the quadrupole ($m = 2$) Fourier component of the jet-related SS 2D  peak.  $A_Q\{\text{2D}\}$ or $B_Q\{\text{2D}\}$  results are insensitive to the SS peak shape on $\eta$ as noted in the present study. The essential model element is the SS peak 1D Gaussian factor on azimuth as noted in Ref.~\cite{bayes}, where 1D Fourier fits without an SS peak element are strongly rejected by a Bayesian-inference analysis.

 \subsection{$\bf y_t$-integral $\bf A_Q\{\text{2D}\}$ systematics} \label{ytintsyst}

The $A_Q\{\text{2D}\}(b,\sqrt{s_{NN}})$ data from Refs.~\cite{davidhq,noelliptic} reveal two interesting features: (a)  Data from two energies are accurately described by the same centrality variation (defined below), (b) the energy dependence of the quadrupole amplitude (in combination with SPS data at 17 GeV~\cite{na49}) scales as $\log(\sqrt{s_{NN}})$. When plotted on $b/b_0$ the $A_Q\{\text{2D}\}$ data reveal a simple variation closely approximating a Gaussian function centered on $b/b_0 = 0.5$.

\begin{figure}[h]
  \includegraphics[width=1.65in,height=1.6in]{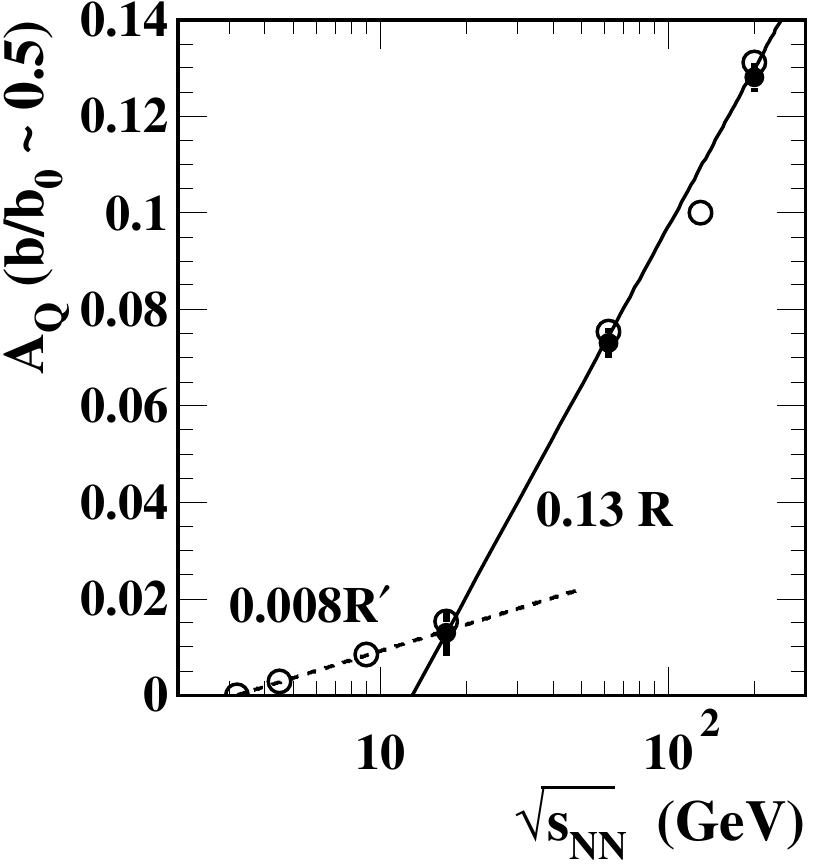}
  \includegraphics[width=1.65in,height=1.6in]{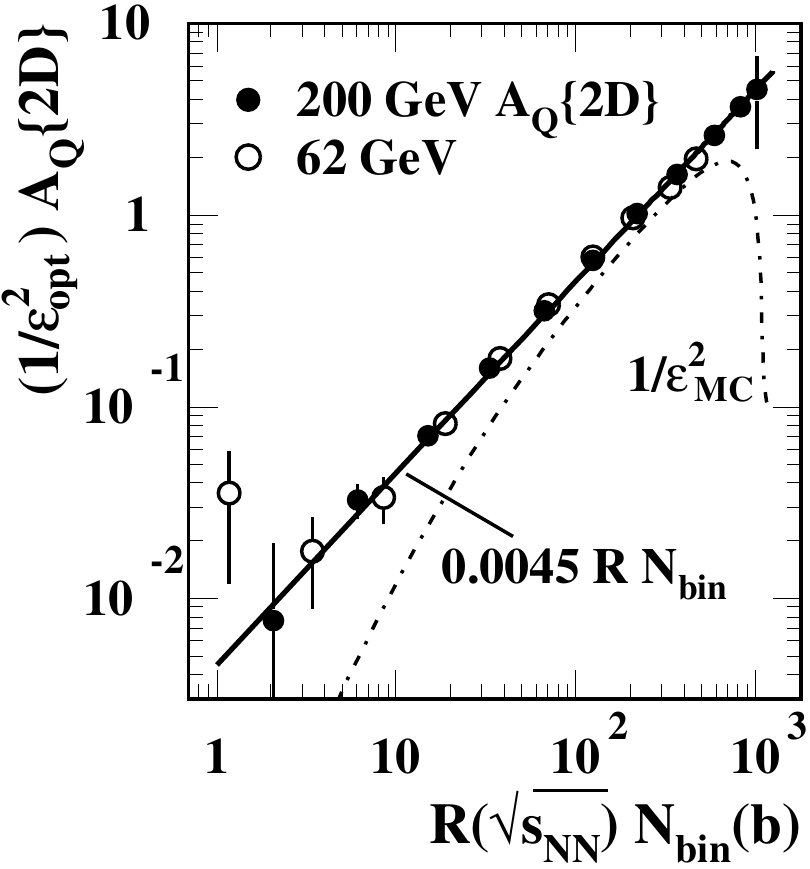}
 \caption{\label{v2eps} 
Left: Collision energy dependence of $v_2$ data converted to per-particle measure $A_Q$. The solid points are from Refs.~\cite{davidhq,noelliptic}  and Ref.~\cite{na49}. They follow a logarithmic trend proportional to $R(\sqrt{s_{NN}})$ (see text).  
The Bevalac-AGS trend below 13 GeV is proportional to $R'(\sqrt{s_{NN}}) = \ln(\sqrt{s_{NN}} / \text{3.2 GeV})$. The open points are taken from Fig.~4 of Ref.~\cite{alicev2}. 
Right: 
The azimuth quadrupole amplitude $A_Q\{\text{2D}\}$ divided by {\em optical} $\epsilon^2_{opt}$ plotted {\em vs} energy-dependent factor $R(\sqrt{s_{NN}})$ times 200 GeV Au-Au binary-collision number $N_{bin}(b)$. The $A_Q\{\text{2D}\}$ data are consistent with simple proportionality [Eq.~(\ref{loglog})] over three decades.
} 
 \end{figure}

Figure~\ref{v2eps} (left panel) summarizes the measured NJ quadrupole energy dependence from Bevalac to highest RHIC energy. $A_Q$ data values at $b/b_0 \approx 0.5$ minimize the relative effects of jet (nonflow) contributions to $A_Q\{\text{method}\}$. 
We observe a major transition in the energy trend of per-particle measure $A_Q$ near 13 GeV, suggesting {\em different physical mechanisms} for the measured NJ quadrupole within the two energy regimes~\cite{gluequad}.
Above 13 GeV the function $R(\sqrt{s_{NN}}) \equiv \log\{\sqrt{s_{NN}} / 13.5\ \text{GeV}\} /\log(200 / 13.5)$ (solid line) describes the energy dependence, with zero intercept at $13.5\pm 0.5$ GeV. 
A similar energy dependence was observed for $\langle p_t \rangle$ fluctuations/correlations attributed to (mini)jets~\cite{edep} consistent with jet-related trends observed recently at the LHC~\cite{ppcms}. 
The rate of increase with energy (line slopes) is {\em six times greater} at higher energies than the Bevalac/AGS trend (dashed curve).

Figure~\ref{v2eps} (right panel) shows $(1/\epsilon_\text{2,opt}^2) A_Q\{\text{2D}\}(b)$  {\em vs}\, $R(\sqrt{s_{NN}}) N_{bin}$, where $N_{bin}$ is the number of \nn\ binary collisions. The vertical-axis variable is motivated to test expectations for the ratio $v_2/\epsilon$: the trend $v_2/ \epsilon \propto S\, dn_{ch}/d\eta $, with $S$ the \aa\ overlap area for more-peripheral \aa\ collisions should transition to a ``hydro limit'' $v_2 / \epsilon \approx $ constant in more-central collisions~\cite{volposk}. 
From the present analysis we conclude that for Au-Au collisions $A_Q\{\text{2D}\}$ data above 13 GeV are described by
\bea \label{loglog}
A_Q\{\text{2D}\}(b,\sqrt{s_{NN}}) &\equiv& \rho_0(b) v_2^2\{\text{2D}\}(b,\sqrt{s_{NN}}) \\ \nonumber
&=& C R(\sqrt{s_{NN}}) N_{bin}(b) \epsilon_{2,opt}^2(b),
\eea
where coefficient $C$ is defined by $1000 C = 4.5\pm 0.2$. 
Equation~(\ref{loglog}) accurately describes measured $y_t$-integral azimuth quadrupole data in heavy ion collisions for all centralities down to \nn\ collisions and all energies above $\sqrt{s_{NN}} \approx 13$ GeV. It represents factorization of energy and centrality dependence for the NJ quadrupole. The 2D quadrupole data are also consistent with $V_2^2\{\text{2D}\} = \rho_0 A_Q\{\text{2D}\} \propto N_{{part}} N_{bin}\epsilon_{2,opt}^2(b)$~\cite{noelliptic}, a  trend that, modulo the IS eccentricity factor, increases much faster than the dijet production rate. The non-zero value $v_2 \approx 0.02$ from Eq.~(\ref{loglog}) extrapolated to \pp\ (\nn) collisions agrees with a \pp\ color-dipole prediction from QCD theory~\cite{boris}.

\subsection{$\bf y_t$-differential $\bf A_Q\{\text{2D}\}$ systematics}

The $B_Q\{\text{2D}\}(y_t,b,\sqrt{s_{NN}})$ data from the present study reveal two interesting features: (a) the quadrupole source boost distribution is independent of \auau\ centrality over a large interval (70\% to 5\%) and (b) the quadrupole spectrum shape is  independent of centrality over the same interval and very different from the SP spectrum shape representing most hadrons. The quadrupole spectrum is much colder (90 MeV vs 145 MeV) and does not change shape above the sharp transition in jet properties as does the SP spectrum (e.g.\ $R_{AA}$ and ``jet quenching''). Those interesting trends are not apparent from the systematics of ratio measure $v_2(p_t)$. Some implications are considered in the next subsection.

\subsection{Physical implications of $\bf A_Q\{\text{2D}\}$ factorization}

The present study combined with previous $y_t$-integral analysis~\cite{davidhq,noelliptic} reveals two factorizations of the NJ quadrupole denoted by $v_2\{\text{2D}\}$: (a) $(b,\sqrt{s_{NN}})$ factorization ($y_t$-integral case) above 13 GeV and (b) $(y_t,b)$ factorization at two energies ($y_t$-differential case). Such factorizations become apparent only in terms of extensive correlation measure $V_2^2$ and with accurate distinction between the NJ quadrupole and other structure, including jet-related SS 2D and AS 1D peaks.


In case (a) ($y_t$-integral case) we can aid interpretation by rearranging Eq.~(\ref{loglog}) to obtain
\bea \label{model2}
2 A_Q  \hspace{-.00in }&=& \hspace{-.00in } C\,R(\sqrt{s_{NN}})\,  \nu(b)\! \left[N_{part}(b)  \epsilon_\text{2,opt}^2(b)\right].
\eea
The LHS per-hadron measure of final-state azimuth quadrupole $A_Q\{\text{2D}\} = \rho_0 v_2^2$ (momentum space) is mathematically analogous to the RHS per-participant IS quadrupole measure within the square brackets (configuration space). The two azimuth correlation measures are simply related by the product of participant path length $\nu(b)$ and energy-dependent factor $R(\sqrt{s_{NN}})$. The quadrupole component of the initial A-A overlap source depends only on impact parameter $b$. 
Thus, {\em the final-state NJ quadrupole of produced hadrons (LHS) is simply proportional to the IS quadrupole of the collision participants (RHS, as determined by $\sqrt{s_{NN}},b$,A) over a large kinematic domain, including \nn\ (\pp) collisions.}

A plot of $A_Q$ vs $1-b/b_0$ in Ref.~\cite{noelliptic} suggests that the centrality dependence in Eq.~(\ref{loglog}) may depend only on the {\em relative} impact parameter $b/b_0$ independent of collision energy. The shape of the quadrupole centrality trend may not depend on the absolute size of the collision system, only on the relative geometry of intersecting spheres independent of atomic number $A$. Further studies with lighter nuclei (e.g., Cu-Cu) may test that hypothesis. 

In case (b) ($y_t$-differential case) we observe that a fixed quadrupole spectrum shape is a universal feature of \auau\ collisions over most of the centrality range (70-5\%), and the inferred source boost distribution is narrow with fixed mean value. Both results contrast strongly with hydro expectations.
In the conventional hydro narrative~\cite{hydro2} (i) almost all hadrons emerge from a monolithic flowing bulk medium, and (ii) flows are driven by pressure gradients corresponding to large IS energy densities in more-central \aa\ collisions. Item (i) implies that quadrupole spectra should be equivalent to SP spectra (and thus cancel in ratio $v_2$), that both phenomena should reflect a broad source boost distribution corresponding to Hubble expansion of  the bulk medium, and there should be a close relation with the systematics of ``jet quenching'' in the medium. Item (ii) implies that flow magnitudes should increase strongly with \aa\ centrality, may be negligible in more-peripheral collisions and should correspond with jet-quenching systematics.

In contrast, measured quadrupole spectrum properties suggest hadronization from a cold boosted source, possibly an expanding cylindrical shell. 
There is no correspondence with the sharp transition in jet properties observed near 50\% fractional cross section, suggesting that the quadrupole phenomenon is not related to jet formation through a dense QCD medium. And the narrow fixed boost distribution independent of \aa\ centrality appears to be incompatible with a Hubble scenario that would describe explosive expansion of a bulk medium, the mean boost increasing with \aa\ centrality~\cite{quadspec}.

\section{Summary} \label{summ}

We have obtained azimuth quadrupole component $v_2\{\text{2D}\}$ data from transverse rapidity $y_t$-differential correlations for 62 and 200 GeV \auau\ collisions. Application of novel analysis methods to 2D angular correlations permits accurate isolation of a nonjet (NJ) quadrupole component with simple systematic properties on $y_t$, \auau\ centrality and collision energy.

Conventional $v_2$ analysis is based on nongraphical numerical methods equivalent to fitting 1D azimuth correlations projected from some pseudorapidity $\eta$ acceptance with a single cosine function.
In the present analysis fits with a multi-element fit model  are applied to 2D angular correlations. The fit model is based on identification of certain geometric features in the 2D data without assumptions about physical mechanisms. In \pp\ and more-peripheral \auau\ collisions the data features are then characterized as jet-related or nonjet by comparisons with theory. Those designations are maintained to central \auau\ collisions, although some physical interpretations may be questioned in more-central collisions.

In this analysis we have identified significant ``nonflow'' bias in published $v_2\{\text{method}\}(y_t,b)$ data, the bias derived mainly from a jet-related SS 2D peak. The bias is accurately  predicted by separately-measured SS peak properties. A variety of strategies has been developed previously in attempts to reduce the nonflow (jet) bias in conventional $v_2$ data, but the results are inconclusive.


The systematics of  $y_t$-differential $v_2\{\text{2D}\}(y_t,b,\sqrt{s_{NN}})$ data from the present study and published $y_t$-integral $v_2\{\text{2D}\}(b,\sqrt{s_{NN}})$ data reveal that the quadrupole power-spectrum amplitude $V_2^2\{\text{2D}\}(y_t,b,\sqrt{s_{NN}})$ derived from those data is fully factorizable. The separate factors on rapidity, centrality and energy are represented by simple functional forms.
In terms of per-particle quadrupole measure $A_Q = \rho_0 v_2^2$ ($\rho_0$ is the single-particle density) the energy dependence is observed to be proportional to $\log(s / s_0)$ ($\sqrt{s_0}  \approx$ 13 GeV) as expected for a QCD process. The centrality dependence is essentially Gaussian on relative impact parameter $b/b_0$. The quadrupole power-spectrum centrality trend is  $V_2^2\{\text{2D}\}(b,\sqrt{s_{NN}}) \propto N_{part} N_{bin}\epsilon_{opt}^2(b)$. The same trends accurately describe data from \pp\ to mid-central \auau\ collisions. A nonzero $v_2$ value for \pp\ collisions derived by extrapolation is consistent with a theory prediction based on an alternative (nonflow) QCD mechanism for the NJ quadrupole.

From the $y_t$-dependence factor {\em quadrupole spectra} can be reconstructed, and a quadrupole {\em source boost} inferred for each collision system. The quadrupole spectrum shape is the same for three hadron species and for all collision systems, and the quadrupole source boost (a single value) is approximately independent of \auau\ centrality.

Our results have implications for hydrodynamic interpretations of \aa\ collisions. The universal quadrupole centrality trend can be contrasted with the trends for jet-related correlations which exhibit a common sharp transition within a small centrality interval, from \nn\ linear superposition in more-peripheral \auau\ collisions to a substantially different dependence in more-central collisions. In contrast, the trend for  $v_2\{\text{2D}\}/\epsilon_{optical}$ remains smooth and slowly varying from \pp\ to more-central \auau\ collisions. If jet production responds to formation of or changes in a dense bulk medium the azimuth quadrupole appears unresponsive to such a medium.

Further implications for hydro models arise from quadrupole spectrum results and quadrupole source-boost trends. The mean source boost does not vary significantly with \auau\ centrality, and the narrow boost distribution is inconsistent with Hubble flow of an expanding bulk medium. The NJ quadrupole and ratio $v_2\{\text{2D}\}/\epsilon$ fall to zero for most-central \auau\ collisions.



\end{document}